\documentclass[12pt, draftclsnofoot, onecolumn]{IEEEtran}
\usepackage{caption}
\usepackage{xcolor}
\usepackage[T1]{fontenc}
\usepackage{cite}
\usepackage[justification=centering]{caption}
\usepackage[linesnumbered,ruled,vlined]{algorithm2e}
\usepackage{amsmath}
\usepackage{epsfig,amssymb,subfigure,bm,dsfont}
\usepackage{algpseudocode}
\usepackage{amsfonts}
\usepackage{array}
\usepackage{footnote}
\usepackage{graphicx}
\usepackage{multicol}
\usepackage{multirow}
\usepackage{subfig}

\newtheorem{theorem}{Theorem}

\newtheorem{definition}{Definition}
\newtheorem{lemma}{Lemma}

\newcommand{\tabincell}[2]{\begin{tabular}{@{}#1@{}}#2\end{tabular}}

\renewcommand{\algorithmicrequire}
{\textbf{Require:}}
\renewcommand{\algorithmicensure}
{\textbf{Ensure:}}
\renewcommand{\algorithmicreturn}
{\textbf{Return}}

\ifCLASSINFOpdf
\else
\fi

\begin{document}

\title{Blockchain Assisted Federated Learning over Wireless Channels: Dynamic Resource Allocation and Client Scheduling}

\author{Xiumei~Deng,~Jun~Li,~Chuan~Ma,~Kang~Wei,~Long~Shi,~Ming~Ding,~Wen~Chen,\\~and~H.~Vincent~Poor
	\thanks{An earlier version of this paper was presented in part at the 2021 IEEE International Conference on Cyber, Physical and Social Computing (CPSCom)\cite{DBLP:conf/ithings/DengLSWWW21} [DOI: 10.1109/iThings-GreenCom-CPSCom-SmartData-Cybermatics53846.2021.00057].}
	\thanks{X. Deng, J. Li, K. Wei and L. Shi are with the School of Electronic and Optical Engineering, Nanjing University of Science and Technology, Nanjing, 210094, China. E-mail: \{xiumeideng, jun.li, kang.wei\}@njust.edu.cn, slong1007@gmail.com.}
	\thanks{C. Ma is with Zhejiang Lab, Hangzhou, China. He is also with Nanjing University of Science and Technology, and Key Laboratory of Computer Network and Information Integration (Southeast University), Ministry of Education. E-mail: chuan.ma@zhejianglab.edu.cn.}
	\thanks{M. Ding is with Data61, CSIRO, Kensington, WA 6152, Australia. E-mail: ming.ding@data61.csiro.au.}
	\thanks{W. Chen is with the Department of Electronics Engineering, Shanghai Jiao Tong University, Shanghai 200240, China. E-mail: wenchen@sjtu.edu.cn.}
	\thanks{H. V. Poor is with Department of Electrical and Computer Engineering, Princeton University, NJ 08544, USA. E-mail: poor@princeton.edu.}}

\IEEEtitleabstractindextext{%
\begin{abstract}
The blockchain technology has been extensively studied to enable distributed and tamper-proof data processing in federated learning (FL). Most existing blockchain assisted FL (BFL) frameworks have employed a third-party blockchain network to decentralize the model aggregation process. However, decentralized model aggregation is vulnerable to pooling and collusion attacks from the third-party blockchain network. Driven by this issue, we propose a novel BFL framework that features the integration of training and mining at the client side. \textcolor{black}{To optimize the learning performance of FL, we propose to maximize the long-term time average (LTA) training data size under a constraint of LTA energy consumption. To this end, we formulate a joint optimization problem of training client selection and resource allocation (i.e., the transmit power and computation frequency at the client side), and solve the long-term mixed integer non-linear programming based on a Lyapunov technique. In particular, the proposed dynamic resource allocation and client scheduling (DRACS) algorithm can achieve a trade-off of [$\mathcal{O}(1/V)$, $\mathcal{O}(\sqrt{V})$] to balance the maximization of the LTA training data size and the minimization of the LTA energy consumption with a control parameter $V$. Our experimental results show that the proposed DRACS algorithm achieves better learning accuracy than benchmark client scheduling strategies with limited time or energy consumption.}
\end{abstract}

\begin{IEEEkeywords}
Federated learning, blockchain, Lyapunov optimization, resource allocation, client scheduling.
\end{IEEEkeywords}}

\maketitle
\IEEEdisplaynontitleabstractindextext

\IEEEpeerreviewmaketitle

\section{Introduction}

\IEEEPARstart{W}{ith} the emergence of the Internet of Things (IoT), very large amounts of data are being generated by smart devices with increasingly powerful computation and sensing capabilities, which motivates the development of Artificial Intelligence (AI) and its applications\cite{DBLP:journals/tits/KumarRD21}. As an emerging AI technology, federated learning (FL) can achieve a global machine learning (ML) model aggregated at a centralized server using the local ML models trained across distributed clients each with a local private dataset. Compared with traditional ML, the clients in FL can collaboratively build a shared model without any raw data exchange, thereby promoting privacy of each client\cite{DBLP:journals/tcss/QuYGZP18}. \textcolor{black}{However, the centralized model aggregation in traditional FL poses the potential threats of inaccurate global model update once the central server is attacked (e.g., data tampering attack) or disruption of FL training when the central server fails due to physical damage (e.g. data transmission failure and model aggregation failure).}

\textcolor{black}{To address the aforementioned issues, blockchain technology can be introduced to FL to eliminate the need of collecting the local models at a central FL server for global model aggregation \cite{li2021blockchain}. Thanks to the advantages of blockchain such as being tamper-proof, anonymity, and traceability, immutable auditability of ML models can be achieved in blockchain to promote trustworthiness in tracking provenance \cite{DBLP:journals/cm/XiongZNWH18}.} Recent works on blockchain assisted FL (BFL) networks have mainly proposed to offload the global model aggregation to a group of distributed servers that form an independent blockchain network\cite{DBLP:journals/tii/LuHDMZ20a,DBLP:conf/bcca/KorkmazKUMOA20,DBLP:journals/tcom/PokhrelC20,DBLP:conf/globecom/OtoumRM20,DBLP:journals/tii/QuPGGX21,feng2021blockchain}. \textcolor{black}{However, model aggregation in BFL networks is vulnerable to pooling and collusion attacks from the third-party blockchain network, where colluding miners with majority control of the network's mining hash rate can manipulate the model aggregation by denying legitimate blocks and creating biased ones \cite{DBLP:journals/cim/MaLSDWHP22,DBLP:journals/corr/abs-2104-01918}.} The possibility of such a pooling attack undermines the core value of blockchain, i.e., the decentralization, harms its security, and degrades the learning performance of FL.

Consensus algorithms in blockchain such as Proof of Work (PoW) rely on intensive computation and energy resources. In reality, PoW is the most prevalent consensus mechanism, which has been widely deployed in mainstream blockchain networks such as Bitcoin and Ethereum. A PoW-enabled blockchain is secured by the miners in the network racing to solve an extremely complicated hash puzzle. As a result, the training latency increases significantly due to the mining process in BFL networks\cite{DBLP:journals/icl/KimPBK20,DBLP:journals/iotj/NguyenDPPLSLNP21}. However, for resource-limited clients, energy and computation constraints may reduce network lifetime and efficiency of training tasks\cite{DBLP:journals/iotj/NguyenDPPLSLNP21,DBLP:journals/corr/abs-2004-04104}, which becomes a crucial bottleneck of BFL. Furthermore, considering the local model transmission over dynamic wireless channels, communication cost also has significant impact on the learning performance of BFL. In view of these challenging issues, the designs of time-efficient and computation-efficient BFL over wireless networks deserve further study.

To address the aforementioned issues, we propose a novel BFL framework that features the integration of training and mining at the client side. For computation-limited and energy-limited BFL networks, we formulate a stochastic optimization problem to optimize the learning performance of FL under either limited training time or limited total energy supply, by maximizing the long-term time average (LTA) training data size under the constraint of LTA energy consumption. With time-varying channel states, we jointly optimize communication, computation, and energy resource allocation as well as client scheduling. Our contributions are summarized as follows:

$\bullet$ By integrating blockchain with FL, we propose a BFL network wherein the role of the client can be either a trainer for its local model training, or a miner for global model aggregation and verification. Specifically, the client transmits it trained model to other clients, performs global aggregation upon receiving others’ models, and competes to mine a block without the intervention of any third-party blockchain network.

$\bullet$ Building upon the proposed framework, we study a wireless BFL network wherein the model communications among different clients are over wireless fading channels. Under the key characteristics of wireless channels, we propose a training client scheduling protocol to meet stringent latency requirement in FL, where the clients with qualified channel conditions are scheduled to train their models in each communication round. Furthermore, we formulate a joint dynamic optimization problem of the training client scheduling and resource allocation (i.e., the transmit and computation power at the client side) under the constraint of LTA energy consumption. The objective of this optimization problem is to maximize the LTA training data size and thereby optimize the learning performance of FL. To this end, we propose a dynamic resource allocation and client scheduling (DRACS) algorithm to obtain a closed-form solution by using the Lyapunov optimization method.

$\bullet$ A performance analysis of the proposed algorithm is conducted to verify its asymptotic optimality. We also characterize a trade-off of [$\mathcal{O}(1/V)$, $\mathcal{O}(\sqrt{V})$] between the LTA training data size and energy consumption with a control parameter $V$. This trade-off indicates that the maximization of the LTA training data size and the minimization of the LTA energy consumption can be balanced by adjusting $V$.

$\bullet$ Our experimental results first show that the proposed algorithm can guarantee the stability of LTA energy consumption. Second, we corroborate the analytical results, and demonstrate the results of energy resource allocation with different sizes of local datasets and LTA energy supply. In addition, the proposed algorithm achieves better learning accuracy than benchmark client scheduling strategies under limited latency constraint.

The remainder of this paper is organized as follows. In Section \ref{sec:related work}, we review the related works and research gaps. In Section \ref{sec:System Model}, we first introduce the wireless BFL network, and then formulate the stochastic optimization problem. Sections \ref{sec:Problem Solution} and \ref{Problem3 Solution} present the DRACS algorithm and problem solution respectively. Section \ref{sec:Trade-Off Analysis} investigates the trade-off between training data size and energy consumption. Then, the experimental results are presented in Section \ref{sec:Experimental Results}. Section \ref{sec:Conclusions} concludes this paper. \textcolor{black}{For ease of reference, Table \ref{tab:SummaryofMainAcronyms} and \ref{tab:SummaryofMainNotations} list the main abbreviations and notations used in this paper, respectively.}
\begin{table}[!t]
	\captionsetup{labelfont={color=black,scriptsize},font={color=black,scriptsize}}
	\caption{List of main abbreviations.}
	\label{tab:SummaryofMainAcronyms}
	\centering
	\scalebox{0.8}{
	{\color{black}\begin{tabular}{c||c|c||c}
	\hline
	\bfseries Abbreviations & \bfseries Descriptions&\bfseries Abbreviations & \bfseries Descriptions\\
	\hline
	FL& Federated learning &AP& Access point\\
	\hline
	BFL& Blockchain assisted federated learning &SVM& Support-vector machine\\
	\hline
	LTA& Long-term time-average &CNN& Convolutional neural network\\
	\hline
	ML& Machine learning &ReLU& Rectified linear unit\\
	\hline
	DRACS&Dynamic resource allocation and client scheduling&PoW& Proof-of-Work\\
	\hline
\end{tabular}}}
\end{table}
\begin{table}[!t]
	\captionsetup{labelfont={color=black,scriptsize},font={color=black,scriptsize}}
	\caption{List of main notations.}
	\label{tab:SummaryofMainNotations}
	\centering
	\scalebox{0.8}{
	{\color{black}\begin{tabular}{c||c|c||c}
		\hline
		\bfseries Notations & \bfseries Descriptions&\bfseries Notations & \bfseries Descriptions\\
		\hline
		$\mathcal{N}$& Index set of the clients&$\tau_n^\text{tra}(t)$& Local model training time of the $n$-th client\\
		\hline
		$\mathcal{D}_n$& Local dataset&$f_n^\text{tra}(t)$& Computation frequency of the $n$-th client for local model training\\
		\hline
		$\boldsymbol{w}_n(t)$& Local model parameters of the $n$-th client&$E_n^\text{tra}(t)$& Energy consumption of the $n$-th client for local model training\\
		\hline
		$\mathcal{T}$& Communication round index&$\tau_n^\text{up}(t)$& Local model transmission time from the $n$-th client to the AP\\
		\hline
		$\tau(t)$& Duration of the $t$-th communication round&$P_n(t)$& Transmit power of the $n$-th client\\
		\hline
		$\boldsymbol{i}(t)$& Training client scheduling vector&$h_n(t)$& Uplink channel power gain from the $n$-th client to the AP\\
		\hline
		$K$& Local epoch&$N_0$& Noise power spectral density\\
		\hline
		$\boldsymbol{W}(t)$& Global model parameters&$D(t)$& Total size of selected training datasets\\
		\hline
		$\beta$& Step size&$f_n^\text{bloc}(t)$& Computation frequency of the $n$-th client for block mining\\
		\hline
		$\gamma_n$& Model size&$E_n^\text{bloc}(t)$& Energy consumption of the $n$-th client for block mining\\
		\hline
		$B$& System bandwidth&$E_n(t)$& Total energy consumption of the $n$-th client\\
		\hline
		$\tau^\text{bloc}(t)$& Block mining time&$E_n^\text{sup}$& LTA energy supply of the $n$-th client\\
		\hline
		$\alpha$& Block generation difficulty&$\overline{E_n}$& LTA energy consumption of the $n$-th client\\
		\hline
	\end{tabular}}}\vspace{-15pt}
\end{table}

\section{Related Works \& Research Gaps}\label{sec:related work}
\textcolor{black}{To eliminate the need of a central FL server for global model aggregation, related works on decentralized FL topologies and systems can be summarized into two main categories, i.e., topology based decentralized FL and blockchain assisted decentralized FL.}

\textit{Topology based decentralized FL:} Related works on topology-based decentralized FL focus on communication protocol design to reduce overall communication complexity and thereby decrease the FL training time. For instance,  \cite{lalitha2018fully,DBLP:journals/corr/abs-1905-06731,DBLP:journals/tii/ChenYZYC22} propose the complete-topology based fully decentralized FL framework, wherein each client can communicate to all other clients in the FL network. To improve the communication efficiency, a ring-topology-based decentralized FL scheme is proposed in \cite{wang2022efficient} to reduce the communication complexity and maintain training performance for 24h on-the-go healthcare services. Further, the work in \cite{mertens2022mgm} proposes a gossip protocol based fully decentralized FL framework to improve the communication and energy efficiency in wireless sensor networks. However, the aforementioned works lack a consensus mechanism to enable a common agreement among the clients about the local model update records in each communication round. As a consequence, malicious clients can conduct untraceable model poisoning attacks by transmitting malicious local model updates to other clients, resulting in the learning performance degradation. Compared with the decentralized FL designs \cite{lalitha2018fully,DBLP:journals/corr/abs-1905-06731,DBLP:journals/tii/ChenYZYC22,wang2022efficient,mertens2022mgm}, our proposed BFL framework enables the curation of local model updates in a tamper-proof and traceable manner, and thereby facilitates the process of tracking malicious clients, although the block mining process can cause additional energy consumption and latency for FL implementation.

\textcolor{black}{\textit{Blockchain assisted decentralized FL:} Related works on blockchain assisted decentralized FL focus on BFL framework design to improve the security and reliability of FL \cite{DBLP:journals/tii/LuHDMZ20a,DBLP:conf/bcca/KorkmazKUMOA20,DBLP:journals/tcom/PokhrelC20,DBLP:conf/globecom/OtoumRM20,DBLP:journals/tii/QuPGGX21,feng2021blockchain}. For instance, the work in \cite{DBLP:conf/globecom/OtoumRM20} proposes a BFL framework based on the Proof-of-Work consensus to ensure end-to-end trustworthiness for autonomous vehicular networking systems, and minimizes the communication and consensus delay by optimizing block arrival rate. The proposed BFL system in \cite{DBLP:journals/tii/QuPGGX21} utilizes the practical Byzantine Fault Tolerance protocol to ensure trustworthy shared training and meet delay requirements in vehicular networks. Moreover, ref. \cite{feng2021blockchain} develops a cross-domain BFL framework, and employs threshold multi-signature smart contracts to provide dynamic authentication services for cross-domain drones. Both the aforementioned works and our work propose a fully decentralized FL by integrating blockchain into FL. However, the previous approaches inevitably introduce a third-party blockchain network to store and verify the local models, which can pose the risk of privacy leakage. Unlike existing studies that rely on a third-party blockchain network for decentralized global model aggregation, our proposed BFL framework features the integration of training and mining at the client side. First, without the intervention of any third-party blockchain network, our proposed BFL framework helps enhance privacy by keeping the local models among the participant clients. Second, by orchestrating local model training and block mining at the client side, our proposed BFL framework helps incentivize the participation of clients to contribute not only their computing power to tamper-resistant model updating in blockchain, but also their local datasets to help provide a robust global model in FL.}

\textit{Resource Allocation in BFL Networks.}
The investigation of communication, computation, and energy resource allocation in the BFL networks has drawn much attention. For example, \cite{DBLP:journals/icl/KimPBK20} proposed a BFL architecture, analysed an end-to-end latency model, and further minimized the FL completion latency by optimizing block generation rate. In \cite{DBLP:journals/iotj/QuGLXYLZ20}, the work proposed a BFL model with decentralized privacy protocols for privacy protection, poisoning attack proof, and high efficiency of block generation, and further derived the optimal block generation rate under the constraints of consensus delay and computation cost. Going forward, introducing the characteristics of wireless communications, dynamic resource allocation of computation and communication resources remains challenging in the BFL networks. To solve this problem, \cite{8998397} proposed an asynchronous BFL scheme to minimize the execution time and maximize the accuracy of model aggregation, while computation and energy resource allocation were not considered in this work. In addition, \cite{DBLP:journals/corr/abs-2004-04104} optimized the training data size, energy consumption for local model training, and the block generation rate to minimize the system latency, energy consumption, and incentive cost while achieving the target accuracy for the global model. However, communication resource allocation was ignored in this work. In this paper, we study a joint dynamic optimization problem of communication and computation resources for the proposed BFL network.

\textcolor{black}{For ease of reference, Table \ref{tab:SummaryofRelatedWorks} summarizes the state-of-the-art works on BFL.}
\begin{table}[!t]
	\captionsetup{labelfont={color=black,scriptsize},font={color=black,scriptsize}}	
	\caption{Summary of the related works on BFL.}
	\label{tab:SummaryofRelatedWorks}
	\centering
	\scalebox{0.8}{
	\color{black}{\begin{tabular}{|c|c|c|c|c|c|c|c|c|c|c|c|}
		\hline
		\bfseries Features & \bfseries \cite{DBLP:journals/tii/LuHDMZ20a}& \bfseries \cite{DBLP:conf/bcca/KorkmazKUMOA20}& \bfseries \cite{DBLP:journals/tcom/PokhrelC20}& \bfseries \cite{DBLP:conf/globecom/OtoumRM20}& \bfseries \cite{DBLP:journals/tii/QuPGGX21}& \bfseries \cite{feng2021blockchain}& \bfseries \cite{DBLP:journals/icl/KimPBK20}& \bfseries \cite{DBLP:journals/corr/abs-2004-04104}& \bfseries \cite{DBLP:journals/iotj/QuGLXYLZ20}& \bfseries \cite{8998397}&\bfseries Our work\\
		\hline
		Third-party blockchain network& $\checkmark$& $\checkmark$& $\checkmark$& $\checkmark$& $\checkmark$& $\checkmark$& $\checkmark$& $\checkmark$& $\checkmark$& $\checkmark$& $\times$ \\
		\hline
		Energy efficiency& $\times$& $\times$& $\times$& $\times$& $\times$& $\checkmark$& $\times$& $\checkmark$& $\times$& $\times$& $\checkmark$ \\
		\hline
		Delay minimization& $\times$& $\times$& $\checkmark$& $\checkmark$& $\times$& $\checkmark$& $\checkmark$& $\checkmark$& $\checkmark$& $\checkmark$& $\checkmark$ \\
		\hline
		Communication resource allocation& $\times$& $\times$& $\times$& $\times$& $\times$& $\times$& $\times$& $\times$& $\times$& $\checkmark$& $\checkmark$ \\
		\hline
		Heterogeneous computational capacity& $\times$& $\times$& $\times$& $\times$& $\times$& $\times$& $\times$& $\times$& $\times$& $\checkmark$& $\checkmark$ \\
		\hline
	\end{tabular}}}\vspace{-15pt}
\end{table}

%

\section{System Model}\label{sec:System Model}
\begin{figure}[!t]
	\centering
	\includegraphics[width=5.5in,angle=0]{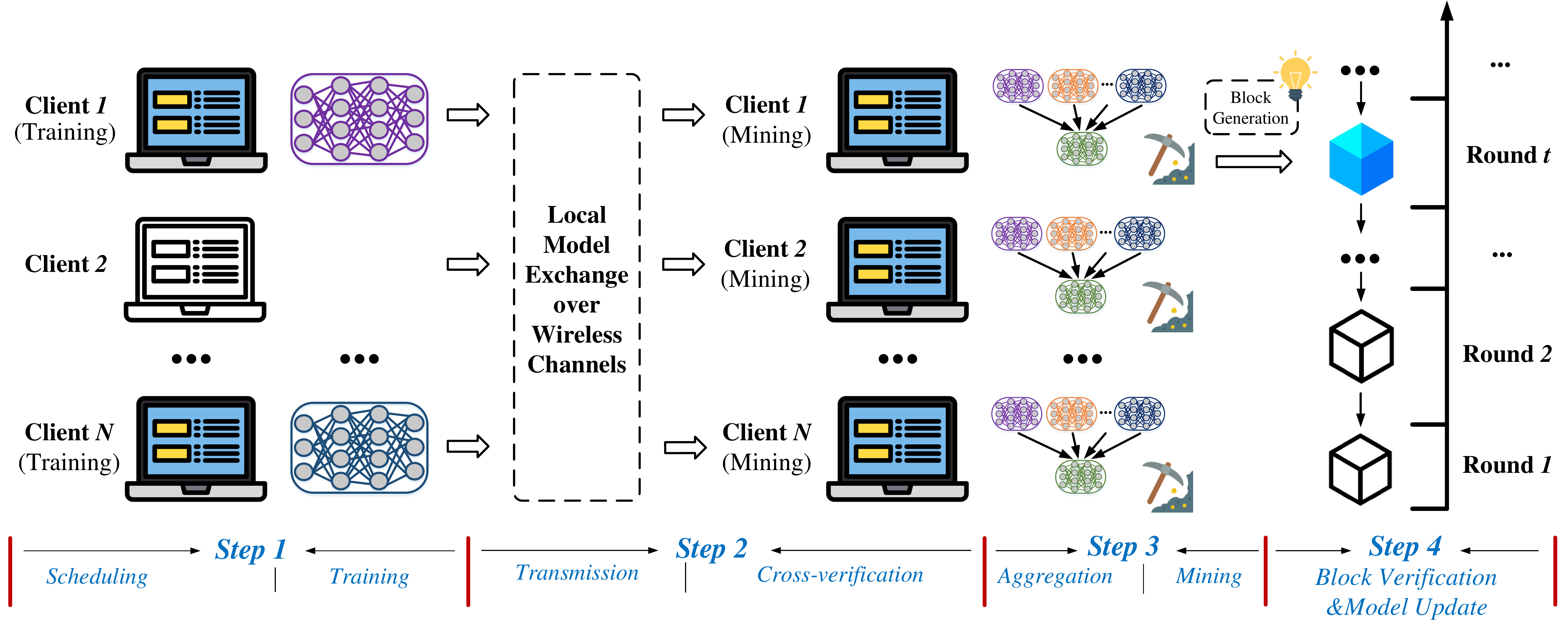}
	\caption{The four-step procedure of proposed wireless BFL in the $t$-th communication round. Each step is detailed in Section \ref{sec:System Model}.}
	\label{fig:system model}\vspace{-0.6cm}
\end{figure}

Fig.\ref{fig:system model} shows a wireless BFL network that consists of $N$ clients. Let $\mathcal{N}=\{1,\ldots,N\}$ denote the index set of the clients. Each client holds a local dataset $\mathcal{D}_n=\{\boldsymbol{x}_{n,m}\in\mathbb{R}^d, y_{n,m}\in\mathbb{R}\}_{m=1}^{D_n}$ with $D_n=\vert\mathcal{D}_n\vert$ sample points, where $\boldsymbol{x}_{n,m}$ is the input vector of the $m$-th sample point at the $n$-th client, and $y_{n,m}$ is the label value of the input. Different from the existing BFL networks \cite{DBLP:journals/tii/QuPGGX21,DBLP:conf/globecom/OtoumRM20,DBLP:journals/tii/LuHDMZ20a,DBLP:conf/bcca/KorkmazKUMOA20,DBLP:journals/tcom/PokhrelC20}, the underlying BFL framework in this paper features the integration of training and mining at the client side, where the roles of each client include local model training, model transmission, and block mining. \textcolor{black}{Considering that FL operation in each communication round is synchronous, each communication round of the proposed BFL operates as follows:}

\textit{Step 1 (Training Client Scheduling and Local Model Training):} A group of clients are selected for local model training at the beginning of each communication round.

\textit{Step 2 (Local Model Transmission and Cross-verification):} The training client encrypts its local model parameters by a unique digital signature, and exchanges the local model parameters with other training clients. Then, all the clients in $\mathcal{N}$ verify the digital signature associated with each set of local model parameters, and store the sets of verified local model parameters locally.

\textcolor{black}{\textit{Step 3 (Global Aggregation and Block Mining):} Each client in $\mathcal{N}$ aggregates the sets of verified local model parameters, and then adds the aggregated model parameters and the sets of verified local model parameters to its candidate block. By following the PoW consensus mechanism, all clients compete to change the nonce and rehash the block header, until the hash value is lower than the target hash set by the block generation difficulty. The first client that finds a valid nonce is the mining winner, and is authorized to add its candidate block to the blockchain.}

\textit{Step 4 (Block Verification and Global Model Update):} The mining winner propagates the new block to the entire network. Upon receiving the new block from the mining winner, each client in $\mathcal{N}$ validates the new block by comparing the model parameters in the new block with its locally stored model parameters. The new block is appended to the blockchain if it can be verified by the majority of clients. Finally, each client updates its local model parameters with the global model parameters in the new block for the next communication round.

\textcolor{black}{In this paper, we orchestrate local model training and block mining at the client side to mitigate the potential threats of privacy leakage and data tampering from malicious mining pools in the third-party blockchain network.}

\textcolor{black}{\textit{Model inversion attack.} The existing BFL designs \cite{DBLP:journals/tii/LuHDMZ20a,DBLP:conf/bcca/KorkmazKUMOA20,DBLP:journals/tcom/PokhrelC20,DBLP:conf/globecom/OtoumRM20,DBLP:journals/tii/QuPGGX21,feng2021blockchain} that rely on a third-party blockchain network for decentralized global model aggregation can pose a potential risk of information leakage. The malicious miners in the third-party blockchain network can conduct differential attacks and model inversion attacks to recover the raw data from the collected local models, which may lead to sensitive information leakage. Without the intervention of any third-party blockchain network, our proposed BFL framework helps enhance privacy by keeping the local models among the participant clients.}

\textcolor{black}{\textit{Data tampering attack.} The resiliency of a third-party blockchain network for global model aggregation can pose the risk of data tampering. Note that in the existing BFL designs \cite{DBLP:journals/tii/LuHDMZ20a,DBLP:conf/bcca/KorkmazKUMOA20,DBLP:journals/tcom/PokhrelC20,DBLP:conf/globecom/OtoumRM20,DBLP:journals/tii/QuPGGX21,feng2021blockchain}, the clients upload the local model updates to their respective associated miners, and then miners broadcast their received local model updates for cross-verification and blocking mining. The malicious miners in the third-party blockchain network can manipulate the model aggregation by tampering with the local model updates uploaded by the associated clients. In addition, colluding miners with majority control of the third-parity blockchain network's mining hash rate or computing power can manipulate the model aggregation by denying legitimate blocks and creating invalid ones, which undermines the core value of blockchain, i.e., the decentralization, harms its security, and degrades the learning performance of FL. Compared with the existing BFL designs, our proposed BFL framework features the integration of training and mining at the client side, which mitigates the potential threats of data tampering and collusion attack from malicious miners in a third-party blockchain network.}

\subsection{Training Client Scheduling and Local Model Training}\label{subsec:Local training model}
The loss function captures the error of the model on the sample points, by calculating the distance between the current output of the global model and the label value of the input. For the $m$-th sample point at the $n$-th client, let us define the loss function as $f(\boldsymbol{w}_n,\boldsymbol{x}_{n,m},y_{n,m})$, where $\boldsymbol{w}_n$ denotes the local model parameters of the $n$-th client. Thus, the loss function on the dataset $\mathcal{D}_n$ is given by $F_n(\boldsymbol{w}_n)=\frac{1}{\vert\mathcal{D}_n\vert}{\sum}_{\{\boldsymbol{x}_{n,m},y_{n,m}\}\in\mathcal{D}_n}f(\boldsymbol{w}_n,\boldsymbol{x}_{n,m},y_{n,m})$. In each communication round, the learning goal of each training client is to minimize $F(\boldsymbol{w}_n)$, i.e., to find
\begin{equation}\label{minimize F}
	\boldsymbol{w}_n^*={\mathop{\arg\min}}_{\boldsymbol{w}_n} F(\boldsymbol{w}_n).
\end{equation}

Due to unaffordable complexity of most ML models, it is  rather challenging to find a closed-form solution to (\ref{minimize F}). Alternatively, (\ref{minimize F}) is solved by using the gradient-descent method as an FL algorithm. Denote the communication round index set by $\mathcal{T}=\{1,\cdots,t,\cdots\}$ and the duration of the $t$-th communication round by $\tau(t)$, respectively. In the $t$-th communication round, each client has its local model parameters $\boldsymbol{w}_n(t)$. Define the training client scheduling vector as $\boldsymbol{i}(t)$ with the $n$-th entry $i_{n}(t)\in \{0,1\}$, $n\in\mathcal{N}$. If $i_{n}(t)=1$, the $n$-th client is selected to train its local model in the $t$-th communication round. Otherwise, the $n$-th client skips its local training. \textcolor{black}{Note that the design of training client scheduling vector $\boldsymbol{i}(t)$ will be given in Section \ref{Problem3 Solution}.} Let $\boldsymbol{w}_n^0(t)$ denote the initial local parameters of the $n$-th client in the $t$-th communication round. At the beginning of the $t$-th communication round, the local parameters for each training client are initialized to the global parameters $\boldsymbol{W}(t-1)$, where the global parameters will be defined in Section \ref{subsec:glo}. After that, the local parameters are updated according to the gradient-descent update rule with respect to the local loss function over a total of $K$ iterations. For each training client, the update rule in the $k$-th iteration is $\boldsymbol{w}_n^k(t)=\boldsymbol{w}_n^{k-1}(t)-\beta \nabla F_n(\boldsymbol{w}_n^{k-1}(t))$, where $\boldsymbol{w}_n^k(t)$ denotes the model parameters of the $n$-th client in the $k$-th iteration and the $t$-th communication round, and $\beta>0$ is the step size.

The local model training time of the $n$-th client in the $t$-th communication round is expressed as 
$\tau_{n}^\text{tra}(t)=\frac{c_nKD_ni_{n}(t)}{f_{n}^\text{tra}(t)}$\cite{9242286},
where $c_n$ is the CPU cycles needed for the $n$-th client to perform the forward-backward propagation algorithm with one sample point, and $f_n^\text{tra}(t)$ is the computation frequency of the $n$-th client for local model training in the $t$-th communication round. Moreover, $f_n^\text{\rm{min}}\leq f_{n}^\text{tra}(t)\leq f_n^\text{\rm{max}}$. The energy consumption of the $n$-th client for local model training in the $t$-th communication round is 
	$E_{n}^\text{tra}(t)=i_{n}(t)v_nc_nKD_n{\left(f_{n}^\text{tra}(t)\right)}^2$, 
where $v_n$ is the effective switched capacitance that depends on the chip architecture.
\subsection{Local Model Transmission}
In the wireless BFL system, an access point (AP) serves as a wireless router for data exchange between different clients\textcolor{black}{\footnote{\textcolor{black}{The clients in the BFL network are expected to forward the data packets via another AP if the current connection fails.}}}  \cite{DBLP:journals/iotj/YuGGG20,DBLP:journals/tvt/ZhanLMWZD20,DBLP:journals/iotj/ZhanHSLN20}. To be specific, the clients transmit the local model parameters and the newly generated blocks to the AP over wireless links, and the AP forwards the local model parameters and new blocks to each client on the network. In addition, we adopt the multiple channel access method of orthogonal frequency-division multiplexing. Consider that the wireless channels are attenuated by independent and identically distributed (i.i.d.) block fading. The channel remains static within each communication round but varies over different rounds. From \cite{DBLP:journals/iotj/YuGGG20}, we model the uplink channel power gain from the $n$-th client to the AP as $h_{n}(t)=h_0\rho_{n}(t)(d_0/d_{n})^{\nu}$. Specifically, $h_0$ is the path loss constant, $d_{n}$ is the distance from the $n$-th client to the AP, $d_0$ is the reference distance, $\rho_{n}(t)$ represents the small-scale fading channel power gain from the $n$-th client to the AP in the $t$-th communication round, and $(d_0/d_{n})^{\nu}$ represents the large-scale path loss with $\nu$ being the path loss factor, which is dominated by the distance. Consider that $\rho^{\text{\rm{min}}}\leq\rho_{n}(t)\leq\rho^{\text{\rm{max}}}$ and the mean value of $\rho_{n}(t)$ is finite, i.e., $\mathbb{E}\{\rho_{n}(t)\} = \overline{\rho_{n}}<\infty$. Thus, the local model transmission time from the $n$-th client to the AP in the $t$-th communication round can be given by\begin{small}\begin{equation}\label{trans time1} \tau_{n}^\text{up}(t)=\frac{\gamma_ni_{n}(t)}{B\log_2\left(1+\frac{P_n(t)h_{n}(t)}{BN_0}\right)},\end{equation}\end{small}where $B$ represents the system bandwidth, $P_n(t)$ denotes the transmit power of the $n$-th client, $N_0$ denotes the noise power spectral density, and $\gamma_n$ is the number of bits that the $n$-th client requires to transmit local model parameters to the AP.

From (\ref{trans time1}), the energy consumption of the $n$-th client for transmitting local model parameters in the $t$-th communication round is given by
$E_n^\text{up}(t)=\frac{P_n(t)\gamma_ni_{n}(t)}{B\log_2\left(1+\frac{P_n(t)h_n(t)}{BN_0}\right)}$.

\subsection{Global Aggregation}\label{subsec:glo}
Upon receiving the models from the training clients, \textcolor{black}{each client in the network performs} the global aggregation by calculating the weighted average of all clients’ local model parameters as
$\boldsymbol{W}(t)=\frac{\sum_{n\in\mathcal{N}} i_n(t)D_n\boldsymbol{w}_n^K(t)}{D(t)}$\cite{9347706},
where $D(t)=\sum_{n\in \mathcal{N}}i_n(t)D_n$ is the total size of selected training datasets in the $t$-th communication round.

\subsection{Block Mining}
After the global model parameters are updated, all the clients in $\mathcal{N}$ complete to mine the new block. From \cite{9079513}, the block mining process under PoW can be formulated as a homogeneous Poisson process. To be specific, the block mining time in each communication round $\tau^\text{bloc}(t)$ is an i.i.d. exponential random variable with the average $\theta(t)=\frac{\alpha}{\sum_{n\in\mathcal{N}} f_n^\text{bloc}(t)}$\cite{9079513,DBLP:journals/iotj/QuGLXYLZ20}, where $f_n^\text{bloc}(t)$ is the computation frequency of the $n$-th client for block mining in the $t$-th communication round, and $\alpha$ is the block generation difficulty. Therefore, the cumulative distribution function of the block mining time in the $t$-th communication round is given by $\text{Pr}(\tau^\text{bloc}(t)<\tau)=1-e^{-\frac{\tau}{\theta(t)}}$.  Thus, with the definition of $p_0=\text{Pr}(\tau^\text{bloc}<\tau)$, the block mining time in the $t$-th communication round is given by
$\tau^\text{bloc}(t)=-\frac{\alpha\ln (1-p_0)}{\sum_{n\in\mathcal{N}} f_n^\text{bloc}(t)}$,
where the new block can be generated as $p_0$ approaches one.

The energy consumption of the $n$-th client for block mining in the $t$-th communication round is expressed as 
$E_n^\text{bloc}(t)=v_n\tau^\text{bloc}(t){\left(f_n^\text{bloc}(t)\right)}^3$.

\subsection{Total Latency and Total Energy Consumption}
\textcolor{black}{Note that the waiting time for each client in the network to collect all of the local models depends on the last client to complete the local model training and transmitting. Considering that the downlink local model transmission time is negligible compared with the overall latency, the total delay of the  each communication round is given by}
\begin{align}\label{latency}
	\textcolor{black}{\tau(t)=\max_{n\in\mathcal{N}}\left\{\tau_{n}^\text{tra}(t)+\tau_n^\text{up}(t)\right\}+\tau^\text{bloc}(t).}
\end{align}

Let $E_n^\text{sup}$ denote the LTA energy supply at the $n$-th client. The total energy consumption of the $n$-th client in the $t$-th communication round is expressed as
\begin{equation}\label{energy111}
	E_{n}(t)=E_{n}^\text{tra}(t)+E_{n}^\text{up}(t)+E_n^\text{bloc}(t),
\end{equation}Thus, the LTA energy consumption of the $n$-th client is given by $\overline{E_{n}}=\lim\limits_{T\to \infty}\frac{\sum_{t=1}^T E_{n}(t)}{\sum_{t=1}^T \tau(t)}$.

\subsection{Problem Formulation}\label{subsec:Problem Formulation}
The training performance of FL in the $t$-th communication round is measured by
\begin{equation}\label{Delta F}
	\Delta F=F(\boldsymbol{W}(t))-F(\boldsymbol{w}^*).
\end{equation}
where $F(\boldsymbol{W}(t))=\sum_{{\boldsymbol{x}_{n,m},y_{n,m}}\in \{\cup_n \mathcal{D}_n\}}\frac{f(\boldsymbol{W}(t),\boldsymbol{x}_{n,m},y_{n,m})}{\vert\{\cup_n \mathcal{D}_n\}\vert}$ is the global loss function in the $t$-th communication round, and $\boldsymbol{w}^*$ is the optimal global model parameters in (\ref{minimize F}). Let $\mathcal{W}$ denote the set of any possible $\boldsymbol{W}(t)$, and assume that $\mathcal{W}$ is convex and bounded. In addition, assume that $f(\boldsymbol{W}(t),\boldsymbol{x}_{n,m},y_{n,m})$ is an $L$-smooth convex loss function on $\boldsymbol{W}(t)$, i.e., $\lVert\nabla_{\boldsymbol{a}} f(\boldsymbol{a},\boldsymbol{x}_{n,m},y_{n,m})-\nabla_{\boldsymbol{b}} f(\boldsymbol{b},\boldsymbol{x}_{n,m},y_{n,m})\lVert\leq L\lVert\boldsymbol{a}-\boldsymbol{b}\lVert$, $\forall\boldsymbol{a},\boldsymbol{b}\in \mathcal{W}$, $\forall\{\boldsymbol{x}_{n,m},y_{n,m}\}\in\{\cup_n \mathcal{D}_n\}$, and the gradient $\nabla_{\boldsymbol{W}(t)} f(\boldsymbol{W}(t),\boldsymbol{x}_{n,m},y_{n,m})$ has a $\sigma^2$ bounded variance for all $\boldsymbol{W}(t)\in \mathcal{W}$, i.e., $\mathbb{E}\{\lVert\nabla_{\boldsymbol{a}} f(\boldsymbol{a},\boldsymbol{x}_{n,m},y_{n,m})-\nabla F(\boldsymbol{a})\lVert^2\}\leq \sigma^2$, $\forall$ $\boldsymbol{a}\in \mathcal{W}$, $\forall$ $\{\boldsymbol{x}_{n,m},y_{n,m}\}\in \{\cup_n \mathcal{D}_n\}$\cite{juditsky2011solving,lan2012optimal}. It can be shown in \cite{xiao2010dual,dekel2012optimal} that for i.i.d. sample points, we have
\begin{small}\begin{equation}\label{bound Delta F}
	\mathbb{E}\{\Delta F\}\leq\frac{2{G}^2L}{D(t)}+\frac{2G\sigma}{\sqrt{D(t)}},
\end{equation}\end{small}
where $G=\sqrt{\max_{\boldsymbol{a},\boldsymbol{b}\in \mathcal{W}}\frac{{\lVert\boldsymbol{a}-\boldsymbol{b}\lVert}^2}{2}}$. Since minimizing (\ref{Delta F}) is intractable, we instead minimize the upper bound of the expectation of (\ref{Delta F}) in (\ref{bound Delta F}) \cite{DBLP:conf/icc/FengWZQP20,DBLP:journals/twc/ChenYSYPC21}. Note that minimizing the right-hand-side of (\ref{bound Delta F}) is equivalent to the maximization of training data size $D(t)$, since the right-hand-side of (\ref{bound Delta F}) has a negative correlation with the training data size $D(t)$. In this case, we maximize the training data size $D(t)$ to optimize the training performance in each communication round.

In addition, due to the limited battery capacity and the charging rate of energy supply, it is crucial to make sure that the LTA energy consumption cannot exceed the LTA energy supply, i.e., $\overline{E_{n}}\leq E_n^\text{sup}$. This guarantees that sufficient energy exists in the batteries for local model training, model transmission, and block mining. Therefore, the goal of this paper is to maximize the LTA training data size under the LTA energy consumption constraint. Let $\boldsymbol{X}(t)=$ [$\boldsymbol{i}(t)$, $\boldsymbol{P}(t)$, $\boldsymbol{f}^\text{tra}(t)$, $\boldsymbol{f}^{\text{bloc}}(t)$]. In this context, we formulate the stochastic optimization problem as
\begin{alignat}{1}\label{P}
	&\quad\quad\quad\quad\quad\quad\quad\quad\quad\quad\quad\textbf{P0}:\;\max_{\boldsymbol{X}(t)}\; \overline{D}=\lim\limits_{T\to \infty}\frac{\sum_{t=1}^TD(t)}{\sum_{t=1}^T\tau(t)}\\
	\text{s.t.}\quad
	&\textbf{C1}: i_n(t)\in \{0,1\},\forall n\in \mathcal{N},t\in \mathcal{T},\quad\quad\quad\quad\;\;\textbf{C2}: P_n^\text{\rm{min}}\leq P_n(t)\leq P_n^\text{\rm{max}}, \forall n\in \mathcal{N},t\in \mathcal{T},\nonumber\\
	&\textbf{C3}: f_n^\text{\rm{min}}\leq f_{n}^\text{tra}(t)\leq f_n^\text{\rm{max}}, \forall n\in \mathcal{N}, t\in \mathcal{T},\quad\textbf{C4}: f_n^\text{\rm{min}}\leq f_{n}^\text{bloc}(t)\leq f_n^\text{\rm{max}}, \forall n\in \mathcal{N}, t\in \mathcal{T},\nonumber\\
	&\textbf{C5}: \overline{E_{n}}<E_n^\text{sup}, \forall n\in \mathcal{N},\nonumber
\end{alignat}
where $\overline{D}$ is the LTA training data size. From (\ref{P}), communication, computation, and energy resources are jointly optimized with LTA energy consumption constraint.
\section{Dynamic Resource Allocation and Client Scheduling Algorithm}\label{sec:Problem Solution}
\begin{figure}[!t]
	\centering
	\includegraphics[width=5.2in,angle=0]{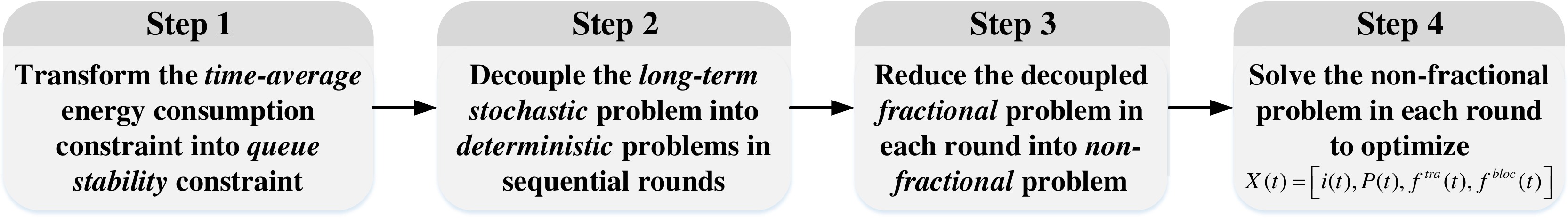}
	\captionsetup{labelfont={color=black,scriptsize},font={color=black,scriptsize}}	
	\caption{Functional workflow of the proposed DRACS approach.}\vspace{-10pt}
	\label{functional flow}
\end{figure}

\begin{small}\begin{algorithm}\small
		\caption{Dynamic Resource Allocation and Client Scheduling}\label{alg::Dynamic Resource Allocation and Client Scheduling Algorithm}
		Initialize:	Virtual queue length $Z_n(t)=0$, tolerance error $\xi>0$\;
		\For{$t=1,2,...$}{
			\algorithmicrequire{ Virtual queue length $\{Z_n(t)\}$ and channel state $\{h_n(t)\}$}\;
			\algorithmicensure{ $\boldsymbol{X}(t)=$ [$\boldsymbol{i}(t)$, $\boldsymbol{P}(t)$, $\boldsymbol{f}^\text{tra}(t)$, $\boldsymbol{f}^{\text{bloc}}(t)$]}\;
			Initialize: $s=0$, lower and upper bound of $\Delta_V(t)$ as $\Delta_t^{\!\text{\rm{min}}}(0)$ and $\Delta_t^{\!\text{\rm{max}}}(0)$\;
			\Repeat{$\vert \inf_{\boldsymbol{X}(t)}U(t)\vert\leq \xi$}{
				Given $\eta=\frac{\Delta_t^{\text{\rm{min}}}(s)+\Delta_t^{\text{\rm{max}}}(s)}{2}$, find $\inf_{\boldsymbol{X}(t)}U(t)$ by solving \textbf{P3}\;
				\lIf{$\inf_{\boldsymbol{X}(t)}U(t)=0$}{Break}
				\lIf{$\inf_{\boldsymbol{X}(t)}U(t)<0$}{$\Delta_t^{\text{\rm{min}}}(s+1)=\Delta_t^{\text{\rm{min}}}(s+1)$, $\Delta_t^{\text{\rm{max}}}(s+1)=\eta$}
				\lIf{$\inf_{\boldsymbol{X}(t)}U(t)>0$}{$\Delta_t^{\text{\rm{min}}}(s+1)=\eta$, $\Delta_t^{\text{\rm{max}}}(s+1)=\Delta_t^{\text{\rm{max}}}(s+1)$}
				Set $s\gets s+1$\;
			}
			Update $\{Z_n(t)\}$ according to (\ref{vitual queues})\;
			\algorithmicreturn{ $\boldsymbol{X}(t)=$ [$\boldsymbol{i}(t)$, $\boldsymbol{P}(t)$, $\boldsymbol{f}^\text{tra}(t)$, $\boldsymbol{f}^{\text{bloc}}(t)$]}
		}
\end{algorithm}\end{small}
In this section, we propose a dynamic resource allocation and client scheduling (DRACS) algorithm shown in \textbf{Algorithm \ref{alg::Dynamic Resource Allocation and Client Scheduling Algorithm}} to solve the stochastic optimization problem. \textcolor{black}{For ease of understanding, Fig. \ref{functional flow} illustrates the functional workflow of the proposed DRACS approach.}
\begin{figure}[!t]
	\centering
	\includegraphics[width=3.5in,angle=0]{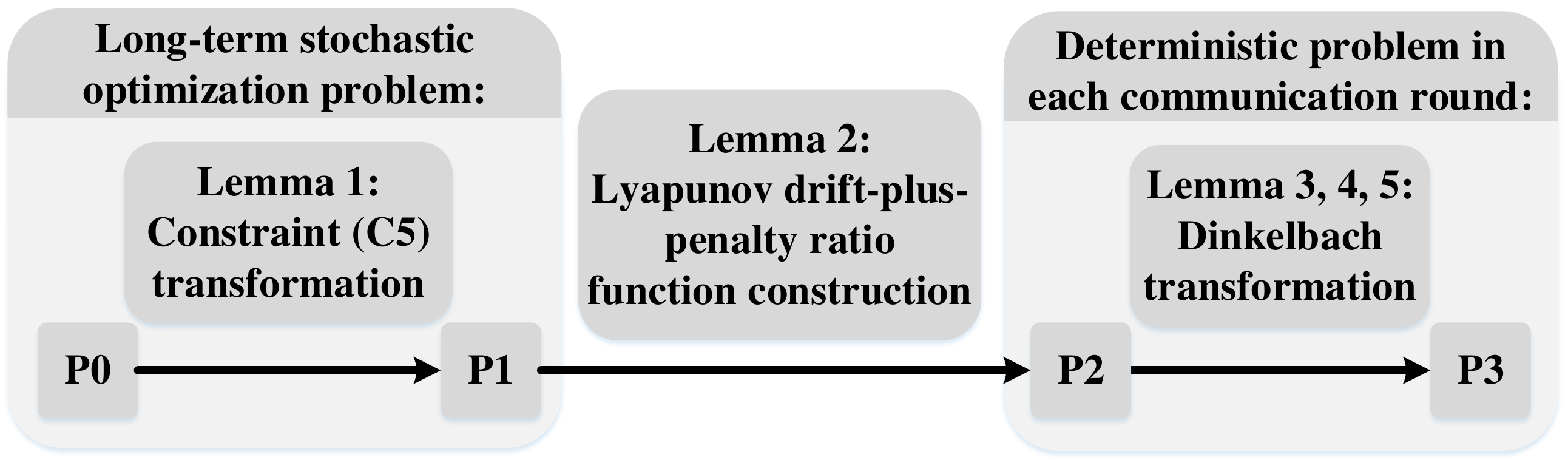}
	\caption{An illustration for the logic flow from problems \textbf{P0} to \textbf{P3}.}\vspace{-0.6cm}
	\label{fig:logic of problem transform}
\end{figure}

With the assistance of the Lyapunov optimization framework, we first transform the time average inequality constraint $\textbf{C5}$ into the queue stability constraint $\tilde{\textbf{C}}\bf{5}$ in \textbf{P1}, and then transform the long-term stochastic problem \textbf{P1} into a deterministic problem \textbf{P2} in each communication round by characterizing the Lyapunov drift-plus-penalty ratio function. By transforming the combinatorial fractional problem \textbf{P2} into the subtractive-form problem \textbf{P3}, the optimal resource allocation and client scheduling policy can be obtained by the Dinkelbach method in an iterative way with a low complexity. In addition, for ease of understanding, Fig. \ref{fig:logic of problem transform} illustrates the main logic flow from \textbf{P0} to \textbf{P3}.

To solve \textbf{P0}, we first transform the time average inequality constraint $\textbf{C5}$ into queue stability constraint. To this end, we define the virtual queues for each client with update equation as
\begin{equation}\label{vitual queues}
	Z_n(t+1)=\max\left\{Z_n(t)+E_{n}(t)-E_n^\text{sup}\tau(t),0\right\}.
\end{equation}
\begin{definition}\label{definition:mean rate stable}
	A discrete time process $Q(t)$ is mean rate stable if $	\lim_{t\to \infty}\frac{\mathbb{E}\{\vert Q(t)\vert\}}{t}=0$ \cite{neely2010stochastic}.
\end{definition}
\begin{lemma}\label{lemma:stability constraint}
	$\rm{\bf{C5}}$ can be satisfied if $Z_n(t)$ is mean rate stable, i.e., $\lim\limits_{t\to \infty}\frac{\mathbb{E}\{\vert Z_n(t)\vert\}}{t}=0$\cite{9242286}.
\end{lemma}
Replacing the LTA energy consumption constraint $\textbf{C5}$ with mean rate stability constraint of $Z_n(t)$, we rewrite \textbf{P0} as
\begin{alignat}{1}\label{P1}
	\textbf{P1}:\max_{\boldsymbol{X}(t)} \;\overline{D}\quad\text{s.t.}\;
	\textbf{C1}\sim\textbf{C4},\;{\tilde{\textbf{C}}\bf{5}}:\lim\limits_{t\to\infty}\frac{\mathbb{E}\{\vert Z_n(t)\vert\}}{t}=0.
\end{alignat}

Now, \textbf{P1} is a standard structure required for Lyapunov optimization. To solve \textbf{P1}, we next formulate the Lyapunov function, characterize the conditional Lyapunov drift, and minimize the Lyapunov drift-plus-penalty ratio function \cite{neely2010stochastic}.
\begin{definition}
	For each $Z_n(t)$, we define the Lyapunov function as $L(t)=\frac{1}{2}{\sum}_{n\in\mathcal{N}}Z_n(t)^2$.
\end{definition}
\begin{definition}
	Let $\boldsymbol{Z}(t)$=$\{Z_n(t)$,$\forall n\in\mathcal{N}\}$ denote the set collecting all virtual queue lengths in the $t$-th round. We define the conditional Lyapunov drift as $\Delta L=\mathbb{E}\{L(t+1)-L(t)\vert\boldsymbol{Z}(t)\}$.
\end{definition}
The conditional Lyapunov drift depends on the resource allocation and client scheduling policy in reaction to time-varying channel state, local computation resources and current virtual queue lengths. Minimizing $\Delta L$ would help stabilize the virtual queues $\boldsymbol{Z}(t)$, which encourages the virtual queues to meet the mean rate stability constraint $\tilde{\textbf{C}}\bf{5}$\cite{neely2010stochastic}. As such, $\textbf{C5}$ can be satisfied according to \textbf{Lemma \ref{lemma:stability constraint}}. However, minimizing $\Delta L$ alone may result in small LTA training data size. To leverage the LTA training data size and energy consumption, we minimize the Lyapunov drift-plus-penalty ratio function instead of minimizing $\Delta L$ alone. In the following, we first characterize an upper bound of $\Delta L$ in \textbf{Lemma \ref{bound of Deltat}}, and derive the Lyapunov drift-plus-penalty ratio function.
\begin{lemma}\label{bound of Deltat}
	Given any virtual queue lengths and any arbitrary $\boldsymbol{X}(t)$, $\Delta L$ is upper bounded by\begin{equation}\label{upper bound of Delta(t)}
		\Delta L\leq H+{\sum}_{n\in \mathcal{N}}\mathbb{E}\left\{Z_n(t)\left(E_{n}(t)-E_n^\text{\rm{sup}}\tau(t)\right)\big|\boldsymbol{Z}(t)\right\},
	\end{equation}where $H=\!\left(\!\frac{P_n^\text{\rm{max}}\gamma_n}{B\log_2(1+{P_n^\text{\rm{min}}h_0\overline{\rho_n}({d_0}/{d_n})^{\nu}}/{(BN_0)})}+c_nv_nD_n{\left(f_n^\text{\rm{max}}\right)}^2-\frac{\alpha v_n{\left(f_n^\text{\rm{max}}\right)}^3\ln(1-p_0)}{\sum_{n\in\mathcal{N}} f_n^\text{\rm{min}}}\right)^2\!\!+\!{\left(E_n^\text{\rm{sup}}\right)}^2\!\left(\frac{\alpha c_nD_n}{f_{n}^\text{\rm{min}}}\!\right.$ $\left.-\!\frac{\alpha\ln (1-p_0)}{\sum_{n\in\mathcal{N}} f_n^\text{\rm{min}}}+\frac{\gamma_n}{B\log_2\!\left(1+{P_n^\text{\rm{min}}h_0\overline{\rho_n}({d_0}/{d_n})^{\nu}}/{(BN_0)}\right)}\right)^2$.
\end{lemma}
\begin{IEEEproof}
	Please see Appendix A.
\end{IEEEproof}

From {\textbf{Lemma \ref{bound of Deltat}}}, it is easy to see that the upper bound of $\Delta L$ can be minimized by minimizing $\mathbb{E}\{\sum_nZ_n(t)E_{n}(t)\}$ and maximizing $\mathbb{E}\{\tau(t)\}$. Based on this, the Lyapunov drift-plus-penalty ratio function can be derived as follows.
\begin{definition}
	Given $V\geq 0$ as a predefined coefficient to tune the trade-off between training data size and virtual queue stability, we define the Lyapunov drift-plus-penalty ratio function as\begin{equation}\label{drift-plus-penalty function}
		\Delta_V=\frac{\mathbb{E}\{-VD(t)+\sum_{n\in \mathcal{N}}Z_n(t)E_{n}(t)\vert\boldsymbol{Z}(t)\}}{\mathbb{E}\{\tau(t)\vert\boldsymbol{Z}(t)\}}.
	\end{equation}
\end{definition}

We minimize the Lyapunov drift-plus-penalty ratio function, where the penalty scaled by the weight $V$ represents how much we emphasize the maximization of training data size. The case for $V>0$ that includes a weighted penalty term corresponds to joint virtual queue stability and training data size maximization.

Note that the Lyapunov drift-plus-penalty ratio function in (\ref{drift-plus-penalty function}) involves conditional expectations. To minimize the Lyapunov drift-plus-penalty ratio function in (\ref{drift-plus-penalty function}), we employ the approach of \textit{Opportunistically Minimizing an Expectation} in \cite[Sect. 1.8]{neely2010stochastic} to generate the optimal policy. That is, in each communication round, we observe the current virtual queue lengths $\boldsymbol{Z}(t)$ and take a control action to minimize
\begin{equation}\label{drift-plus-penalty function'}
	\Delta_V(t)=\frac{-VD(t)+\sum_{n\in \mathcal{N}}Z_n(t)E_{n}(t)}{\tau(t)}.
\end{equation}

Thus, the main idea of the DRACS algorithm is to minimize the Lyapunov drift-plus-penalty ratio function $\Delta_V(t)$ in (\ref{drift-plus-penalty function'}) under any arbitrary positive $V$ in each round, which is written as
\begin{alignat}{1}\label{P2}
	&\textbf{P2}:{\min}_{\boldsymbol{X}(t)} \;\Delta_V(t)\quad\\&\text{s.t.}\;
	\textbf{C1}\sim\textbf{C4}.\nonumber
\end{alignat}
Now we have transformed the long-term stochastic problem in \textbf{P1} into the one-shot static optimization problem in \textbf{P2} in each communication round.

Next, we use the Dinkelbach method to solve the challenging fractional problem \textbf{P2} based on \textbf{Lemmas \ref{Delta_V^*}} and \textbf{\ref{Delta_V min max}}.
\begin{lemma}\label{Delta_V^*}
	Define $\Theta(t)$ as the infimum of $\Delta_V(t)$, i.e., $\Theta(t)= \inf_{\boldsymbol{X}(t)}\left[ \frac{-VD(t)+\sum_{n\in \mathcal{N}}Z_n(t)E_{n}(t)}{\tau(t)}\right]$. Let $U(t)=-VD(t)+{\sum}_{n\in\mathcal{N}}Z_n(t)E_{n}(t)-\eta\tau(t)$. We have ${\inf}_{\boldsymbol{X}(t)}U(t)<0$ if $\eta>\Theta(t)$, and ${\inf}_{\boldsymbol{X}(t)}U(t)>0$ if $\eta<\Theta(t)$.
\end{lemma}\begin{IEEEproof}
	Please see Appendix B.
\end{IEEEproof}
\begin{lemma}\label{Delta_V min max}
	Given any virtual queue lengths $\boldsymbol{Z}(t)$ and any arbitrary $\boldsymbol{X}(t)$, $\Delta_V(t)$ is lower and upper bounded by\vspace{-10pt}
	\begin{small}
		\begin{alignat}{1}
			\Delta_t^{\text{\rm{min}}}\!=\!-\frac{1}{\tau^\text{\rm{max}}}\!\left(V\!\sum_{n\in\mathcal{N}}D_n+\alpha\ln(1-p_0)\!\sum_{n\in\mathcal{N}}\!\frac{v_nZ_n(t){(f_n^\text{\rm{min}})}^3}{{\sum}_{n\in\mathcal{N}} f_n^\text{\rm{max}}}\right),\vspace{-10pt}
	\end{alignat}\end{small}and\vspace{-10pt}
	\begin{small}\begin{alignat}{1}
			\Delta_t^{\text{\rm{max}}}\!&=\!\frac{1}{\tau^\text{\rm{min}}}\!\!\sum_{n\in\mathcal{N}}\!\!Z_n(t)\!\left(\!v_nc_nKD_n{(f_n^\text{\rm{max}})}^2\!-\!\frac{v_n\alpha{(f_n^\text{\rm{max}})}^3\!\ln (1-p_0)}{\sum_{n\in\mathcal{N}} f_n^\text{\rm{min}}}\!\right.\left.+\frac{P_n^\text{\rm{max}}\gamma_n}{B\log_2\left(1+\frac{P_n^\text{\rm{min}}h_0\rho_n^\text{\rm{min}}(d_0/d_n)^{\nu}}{(BN_0)}\right)}\right),\vspace{-10pt}
	\end{alignat}\end{small}where $\tau^\text{\rm{min}}$ and $\tau^\text{\rm{max}}$ are the lower and upper bound of the duration of each communication round, i.e., $\tau^{\text{\rm{min}}}=\max\limits_{n\in\mathcal{N}}\left\{\frac{c_nKD_n}{f_{n}^\text{\rm{max}}}+\frac{\gamma_n}{B\log_2(1+{P_n^\text{\rm{max}}h_0\rho_n^\text{\rm{max}}({d_0}/{d_n})^{\nu}}/{(BN_0)})}\right\}-\frac{\alpha\ln (1-p_0)}{{\sum}_{n\in\mathcal{N}} f_n^\text{\rm{max}}}$, and $\tau^\text{\rm{max}}\!=\!\max\limits_{n\in\mathcal{N}}\left\{\frac{\gamma_n}{B\log_2(1+{P_n^\text{\rm{min}}h_0\rho_n^\text{\rm{min}}({d_0}/{d_n})^{\nu}}/{(BN_0)})}\right.$ $\left.+\frac{c_nKD_n}{f_{n}^\text{\rm{min}}}\right\}-\frac{\alpha\ln (1-p_0)}{{\sum}_{n\in\mathcal{N}} f_n^\text{\rm{min}}}$.
\end{lemma}

The main idea of Dinkelbach method is to solve $\inf_{\boldsymbol{X}(t)}$ $U(t)=0$. That is, we define $\eta=\frac{\Delta_t^{\!\text{\rm{min}}}+\Delta_t^{\!\text{\rm{max}}}}{2}$ and compute the value of $\inf_{\boldsymbol{X}(t)}U(t)$ by solving (see line 7 of \textbf{\textbf{Algorithm \ref{alg::Dynamic Resource Allocation and Client Scheduling Algorithm}}})\begin{alignat}{1}\label{P3}
	&\textbf{P3}:{\min}_{\boldsymbol{X}(t)}\; U(t)\\&\text{s.t.}\;	\textbf{C1}\sim\textbf{C4}.\nonumber
\end{alignat}
In each iteration, if $\inf_{\boldsymbol{X}(t)}U(t)<0$, we have $\eta>\Theta(t)$. Then, we refine the upper bound of $\Delta_V(t)$ as $\Delta_t^{\!\text{\rm{max}}}=\eta$ (see line 10 of \textbf{\textbf{Algorithm \ref{alg::Dynamic Resource Allocation and Client Scheduling Algorithm}}}). Otherwise, we have $\eta<\Theta(t)$, and the lower bound of $\Delta_V(t)$ is refined as $\Delta_t^{\!\text{\rm{min}}}=\eta$ (see line 12 of \textbf{\textbf{Algorithm \ref{alg::Dynamic Resource Allocation and Client Scheduling Algorithm}}}). In this way, the distance between the upper and lower bound of $\Delta_V(t)$ can be reduced to half its original value in each iteration. As such the optimal value of $\Delta_V(t)$ can be approached exponentially fast. At this point, the intractable stochastic optimization problem in \textbf{P0} is transformed into a sequence of deterministic combinatorial problems in \textbf{P3} in each round, which leads to the asymptotically optimal solution. For further details and the proof of convergence, please refer to \cite{dinkelbach1967nonlinear}.

\textcolor{black}{As shown in \textbf{Algorithm 1}, our proposed DRACS algorithm is performed at the client side to optimize the training client scheduling and resource allocation in each communication round. To be specific, all clients in the BFL network exchange virtual queue length and channel state information (see line 3 in \textbf{Algorithm 1}) with each other at the beginning of each communication round. Based on the collected virtual queue lengths and channel states, our proposed DRACS algorithm can be performed at each client to optimize the training client scheduling vector $\boldsymbol{i}(t)$, transmit power $\boldsymbol{P}(t)$, and computation frequency for local training $\boldsymbol{f}^\text{tra}(t)$ and block mining $\boldsymbol{f}^\text{bloc}(t)$ in each communication round (see line 14 in \textbf{Algorithm 1}).}

\section{Optimal Solution for the Sequence of Combinatorial Problems}\label{Problem3 Solution}
In this section, we solve the sequence of deterministic combinatorial problems \textbf{P3} in each communication round. By exploiting the dependence among $\boldsymbol{i}(t)$, $\boldsymbol{P}(t)$, $\boldsymbol{f}^\text{tra}(t)$, $\boldsymbol{f}^{\text{bloc}}(t)$ in the objective function of \textbf{P3},
\textcolor{black}{we first decouple the joint optimization problem into the following two sub-problems, and solve the sub-problems, respectively.}

\subsection{Optimal Computation Frequency For Block Mining}
The computation frequency for block mining $\boldsymbol{f}^\text{bloc}(t)$ of \textbf{P3} can be separately optimized by
\begin{small}\begin{alignat}{1}\label{g2}
		&\min_{\boldsymbol{f}^\text{bloc}(t)} h(\boldsymbol{f}^\text{bloc}(t))=\frac{1}{{\sum}_{n\in\mathcal{N}}f_n^\text{bloc}(t)}\Big(-{\sum}_{n\in\mathcal{N}}Z_n(t)v_n\alpha\ln(1-p_0)\left(f_n^\text{bloc}(t)\right)^3+\eta\alpha\ln(1-p_0)\Big)\\&\text{s.t.}\;\textbf{C4}.\nonumber
\end{alignat}\end{small}To solve the fractional problem in (\ref{g2}), we next derive the lower and upper bounds of $h(\boldsymbol{f}^\text{bloc}(t))$, and solve the optimal computation frequency for block mining $\boldsymbol{f}^\text{bloc}(t)$ by Dinkelbach method.

Given any virtual queue length $\boldsymbol{Z}(t)$, $h(\boldsymbol{f}^\text{bloc}(t))$ is lower and upper bounded by $h^\text{\rm{min}}$ and $h^\text{\rm{max}}$. Recall that $f_n^\text{\rm{min}}\leq f_{n}^\text{bloc}(t)\leq f_n^\text{\rm{max}}$, it can be derived that, if $\eta\ge 0$, $h^\text{\rm{min}}=\frac{\eta\alpha\ln(1-p_0)}{\sum_{n\in\mathcal{N}}f_n^\text{\rm{min}}}+\frac{-\alpha\ln(1-p_0)\sum_{n\in\mathcal{N}}Z_n(t)v_n{(f_n^\text{\rm{min}})}^3}{\sum_{n\in\mathcal{N}}f_n^\text{\rm{max}}}$, and $h^\text{\rm{max}}=\frac{-\alpha\ln(1-p_0){\sum}_{n\in\mathcal{N}}Z_n(t)v_n{(f_n^\text{\rm{max}})}^3}{\sum_{n\in\mathcal{N}}f_n^\text{\rm{min}}}+\frac{\eta\alpha\ln(1-p_0)}{\sum_{n\in\mathcal{N}}f_n^\text{\rm{max}}}$; if $\eta < 0$, $h^\text{\rm{min}}=\frac{-\alpha\ln(1-p_0){\sum}_{n\in\mathcal{N}}Z_n(t)v_n{(f_n^\text{\rm{min}})}^3}{\sum_{n\in\mathcal{N}}f_n^\text{\rm{max}}}+\frac{\eta\alpha\ln(1-p_0)}{\sum_{n\in\mathcal{N}}f_n^\text{\rm{max}}}$, and $h^\text{\rm{max}}=\frac{-\alpha\ln(1-p_0){\sum}_{n\in\mathcal{N}}Z_n(t)v_n{(f_n^\text{\rm{max}})}^3}{\sum_{n\in\mathcal{N}}f_n^\text{\rm{min}}}+\frac{\eta\alpha\ln(1-p_0)}{\sum_{n\in\mathcal{N}}f_n^\text{\rm{min}}}$. Utilizing the lower and upper bounds of $h(\boldsymbol{f}^\text{bloc}(t))$, the problem in (\ref{g2}) can be solved by Dinkelbach method. Let $\mu=\frac{1}{2}(h^\text{\rm{min}}+h^\text{\rm{max}})$ in the first iteration of the Dinkelbach method, \textcolor{black}{the non-linear fractional programming problem in (\ref{g2}) can be transformed into a non-fractional programming problem as follows.}
\begin{small}\begin{alignat}{1}\label{g2'}
		&\min_{\boldsymbol{f}^\text{bloc}(t)} h'(\boldsymbol{f}^\text{bloc}(t))=-{\sum}_{n\in\mathcal{N}}Z_n(t)v_n\alpha\left(f_n^\text{bloc}(t)\right)^3\ln(1-p_0)+\eta\alpha\ln(1-p_0)-\mu{\sum}_{n\in\mathcal{N}}f_n^\text{bloc}(t)\\&\text{s.t.}\;\textbf{C4}.\nonumber
\end{alignat}\end{small}Notably, (\ref{g2'}) is a continuous derivable function. The optimal computation frequency for block mining can be derived as\vspace{-5pt}
\begin{small}\begin{equation}\label{solution of h}
		\begin{split}
			f_n^\text{bloc*}(t)=
			\begin{cases}
				&f_n^\text{\rm{min}},\quad\mbox{if}~\;\sqrt{{-\mu}/{(3Z_n(t)v_n\alpha\ln(1-p_0))}}\leq f_n^\text{\rm{min}},\\
				&f_n^\text{\rm{max}},\quad\mbox{if}~\;\sqrt{{-\mu}/{(3Z_n(t)v_n\alpha\ln(1-p_0))}}\ge f_n^\text{\rm{max}},\\
				&\sqrt{{-\mu}/{(3Z_n(t)v_n\alpha\ln(1-p_0))}},\quad\mbox{otherwise}.
			\end{cases}
		\end{split}
\end{equation}\end{small}Recalling that $Z_n(t+1)=\max\{Z_n(t)+E_{n}(t)-E_n^\text{sup}\tau(t),0\}$ in (\ref{vitual queues}), the virtual queue backlog $Z_n(t)$ increases when the LTA energy consumption exceeds the LTA energy supply, i.e., $\lim\limits_{T\to \infty}\frac{\sum_{t=1}^T E_{n}(t)}{\sum_{t=1}^T \tau(t)}\ge E_n^\text{sup}$. From the optimal policy of computation frequency for block mining in (\ref{solution of h}), the $n$-th client reduces the computation frequency for block mining when there exit a large amount of virtual queue backlogs $Z_n(t)$, such that sufficient energy can be kept in the batteries for local model training, model transmission, and block mining in the coming rounds.

\subsection{Optimal Client Scheduling Vector, Transmit Power, Computation Frequency for Local Model Training}\label{Sec:VB}
The optimal client scheduling vector $\boldsymbol{i}(t)$, transmit power $\boldsymbol{P}(t)$, and computation frequency for local model training $\boldsymbol{f}^\text{tra}(t)$ of \textbf{P3} can be separately optimized by
\begin{small}\begin{alignat}{1}\label{g_1}
		\min_{\boldsymbol{P}(t), \boldsymbol{f}^\text{tra}(t), \boldsymbol{i}(t)} &g(\boldsymbol{P}(t), \boldsymbol{f}^\text{tra}(t), \boldsymbol{i}(t))\!=\!-V\!\!\sum_{n\in\mathcal{N}}\!\! D_n i_n(t)\!+\!\!\!\sum_{n\in\mathcal{N}}\! Z_n(t)\!\!\left(\!i_{n}(t)v_nc_nKD_n{(f_{n}^\text{tra}(t))}^2\!\!+\!\frac{P_n(t)i_{n}(t)\gamma_n}{B\log_2\!\big(1\!+\!\frac{P_n(t)h_n(t)}{BN_0}\big)}\!\right)\nonumber\\&-\eta\max_{n\in\mathcal{N}}\Big\{\left.{i_{n}(t)c_nKD_n}/{f_{n}^\text{tra}(t)}+{i_{n}(t)\gamma_n}/{\Big(B\log_2\Big(1+{P_n(t)h_n(t)}/{(BN_0)}\Big)\Big)}\right\}\\
		\text{s.t.}\quad
		&\textbf{C1}\sim\textbf{C3}.\nonumber
\end{alignat}\end{small}To solve this problem, we decompose (\ref{g_1}) into three sub-problems (\ref{g_1_1}), (\ref{g_1_2}), and (\ref{g_1_3}), solve each sub-problem by convex optimization methods while holding the remaining variables fixed, and optimize (\ref{g_1}) by applying the block coordinate decent method. \textcolor{black}{Note that the sub-problems are exactly solved with optimality in each iteration in order to guarantee the convergence to at least a local optimum\cite{DBLP:journals/tvt/FuMWY21,DBLP:journals/tvt/HuaWWDHY19,DBLP:journals/twc/WuZZ18}.}
\subsubsection{Optimal Client Scheduling Vector}
Given the optimized computation frequency for local model training $\boldsymbol{f}^\text{tra}(t)$ and transmit power $\boldsymbol{P}(t)$, we can rewrite (\ref{g_1}) as\begin{small}\begin{alignat}{1}\label{g_1_1}
		&\min_{\boldsymbol{i}(t)}\; g_1(\boldsymbol{i}(t))=\!\!\sum_{n\in\mathcal{N}}\!i_n(t)\left\{\!\!-VD_n\!+\!\frac{Z_n(t)P_n(t)\gamma_n}{B\log_2\big(1\!+\!\frac{P_n(t)h_n(t)}{BN_0}\big)}+v_nc_nKD_nZ_n(t){\left(f_{n}^\text{tra}(t)\right)}^2\right\}-\eta\max_{n\in\mathcal{N}}\left\{\!\frac{c_nKD_ni_{n}(t)}{f_{n}^\text{tra}(t)}\!\right.\nonumber\\&\left.\quad\quad\quad\quad\;\,\quad\quad+{\gamma_ni_{n}(t)}/{\Big(B\log_2\Big(1+{P_n(t)h_n(t)}/{(BN_0)}\Big)\Big)}\right\}\\
		&\text{s.t.}\quad\textbf{C1}.\nonumber
\end{alignat}\end{small}\textcolor{black}{We employ the approach of case analysis to solve the non-linear integer programming problem (\ref{g_1_1}). That is, we first split the problem (\ref{g_1_1}) into $N$ disjoint cases, and then solve each case separately. Note that the optimal solution of the original problem (\ref{g_1_1}) belongs to the union of the solutions to each case. Thus, the optimal client scheduling vector $\boldsymbol{i}(t)$ is obtained by comparing the solutions to each case. Please see the detailed solution of (\ref{g_1_1}) in Appendix C.}

\subsubsection{Optimal Computation Frequency for Local Model Training}
Given the optimized client scheduling vector variables $\boldsymbol{i}(t)$ and transmit power $\boldsymbol{P}(t)$, we can rewrite (\ref{g_1}) as
\begin{small}\begin{alignat}{1}\label{g_1_2}
		\min_{\boldsymbol{f}^\text{tra}(t)}\; &g_2\left(\boldsymbol{f}^\text{tra}(t)\right)\!=\!\!\sum_{n\in\mathcal{N}}\!i_n(t)Z_n(t)v_nc_nKD_n{\left(f_{n}^\text{tra}(t)\right)}^2\!-\!\eta\max_{n\in\mathcal{N}}\left\{\frac{c_nKD_ni_{n}(t)}{f_{n}^\text{tra}(t)}\!+\!\frac{\gamma_ni_{n}(t)}{B\log_2\big(1+\frac{P_n(t)h_n(t)}{BN_0}\big)}\right\}\\
		\text{s.t.}\quad&\textbf{C3}.\nonumber
\end{alignat}\end{small}\textcolor{black}{Based on the optimized client scheduling vector $\boldsymbol{i}(t)$, it can be derived that $R=\sum_{n\in\mathcal{N}}i_n(t)$ clients are selected for local training and transmitting. Using the approach of case analysis, we first split the min-max optimization problem of (\ref{g_1_2}) into $R$ disjoint cases, and solve each case separately. The optimal computation frequency for local model training $\boldsymbol{f}^\text{tra}(t)$ is obtained by comparing the solutions to each case. Please see the detailed solution of (\ref{g_1_2}) in Appendix D.}

\subsubsection{Optimal Transmit Power}
Given the optimized client scheduling vector variables $\boldsymbol{i}(t)$ and computation frequency for local model training $\boldsymbol{f}^\text{tra}(t)$, we can rewrite (\ref{g_1}) as\begin{small}
	\begin{alignat}{1}\label{g_1_3}
		\min_{\boldsymbol{P}(t)}\; &g_{3}(\boldsymbol{P}(t))=\sum_{n\in\mathcal{N}}\frac{i_n(t)Z_n(t)P_n(t)\gamma_n}{B\log_2\left(1\!+\!\frac{P_n(t)h_n(t)}{BN_0}\right)}-\eta\max_{n\in\mathcal{N}}\left\{\frac{c_nKD_n}{f_{n}^\text{tra}(t)}\!i_{n}(t)+\frac{\gamma_ni_{n}(t)}{B\log_2\left(1\!+\!\frac{P_n(t)h_n(t)}{BN_0}\right)}\right\}.\\
		\text{s.t.}\quad
		&\textbf{C2}.\nonumber
\end{alignat}\end{small}To solve (\ref{g_1_3}), we first optimize transmit power $\boldsymbol{P}(t)$ based on Dinkelbach method in the $R$ cases, and compare the value of $g_{3}(\boldsymbol{P}(t))$ among different cases. The detailed solution to (\ref{g_1_3}) is omitted here, since it largely follows that to (\ref{g_1_2}). 

\section{Performance and Complexity Analysis}\label{sec:Trade-Off Analysis}
\subsection{Performance Analysis}
In this subsection, we will provide the performance analysis of the proposed algorithm to verify asymptotic optimality, and characterizes the trade-off between training data size and energy consumption.

To facilitate the analysis, we first define a $C$-additive approximation\cite{neely2010stochastic} of the DRACS algorithm in (\ref{C-additive approximation}), and derive the trade-off between the LTA training data size and energy consumption in \textbf{Theorem \ref{performance analysis}}. Fix a constant $C\ge0$, using a $C$-additive approximation of the DRACS algorithm in each communication round, we have\begin{small}\begin{alignat}{1}\label{C-additive approximation}
		\Delta_V\leq C\!+\!\inf_{\boldsymbol{X}(t)}\left[\frac{\mathbb{E}\{-VD(t)\!+\!\sum_{n\in \mathcal{N}}Z_n(t)E_{n}(t)\vert\boldsymbol{Z}(t)\}}{\mathbb{E}\{\tau(t)\vert\boldsymbol{Z}(t)\}}\right].\!\!
\end{alignat}\end{small}Let $\phi^\text{opt}$ denote the maximum utility of \textbf{P0} over all control policies. Let $\boldsymbol{X}^*(t)$ denote the actions under the optimal policy of \textbf{P2}, and $\phi(\hat{\boldsymbol{X}^*})$ represent the corresponding maximum utility, where $\hat{\boldsymbol{X}^*}=[\boldsymbol{X}^*(1),...,\boldsymbol{X}^*(T)]$. Note that the optimal solution of \textbf{P3} is the asymptotically optimal solution of \textbf{P2}. \textbf{Theorem \ref{performance analysis}} below verifies that $\phi(\hat{\boldsymbol{X}^*})$ of \textbf{P2} converges to $\phi^\text{opt}$ of \textbf{P0} as $V$ increases, and the LTA energy consumption of each client decreases and finally converges to the LTA energy supply $E_n^\text{sup}$ as $V$ decreases and $T$ increases. It is also shown that there exists an [$\mathcal{O}(1/V)$, $\mathcal{O}(\sqrt{V})$] trade-off between the LTA training data size and energy consumption with a control parameter $V$. With a control parameter $V$ to tune the [$\mathcal{O}(1/V)$, $\mathcal{O}(\sqrt{V})$] trade-off between the maximization of the LTA training data size and the minimization of the LTA energy consumption, a large value of $V$ can be utilized to increase the LTA training data size and thereby speed up the BFL process for delay-sensitive applications, and a small value of $V$ can be utilized for energy-sensitive and delay-tolerant applications.	
\begin{theorem}\label{performance analysis}
	With the optimal policy of \rm{\textbf{P2}} implemented as a $C$-additive approximation in each communication round, and note that $\mathbb{E}\{Z(0)\}<\infty$, there exists
	\begin{small}\begin{equation}\label{equ2:theorem1}
			\phi^\text{\rm{opt}}-\phi(\hat{\boldsymbol{X}^*}) \leq \frac{H/\tau^\text{\rm{min}}+C}{V},
	\end{equation}\end{small}and
	\begin{small}\begin{equation}\label{equ3:theorem1}
			\overline{E_{n}}(T) \leq E_n^\text{\rm{sup}}+\frac{1}{\tau^\text{\rm{min}}}\sqrt{\frac{G_1+VG_2}{T}+\frac{\sum_{n\in\mathcal{N}}\mathbb\{({Z_n(0))}^2\}}{T^2}},
	\end{equation}\end{small}where $\overline{E_{n}}(T)=\frac{\sum_{t=1}^TE_{n}(t)}{\sum_{t=1}^T\tau(t)}$, $G_1=2(H+\tau^\text{\rm{max}})$, and $G_2=2(\tau^\text{\rm{max}}\phi^\text{\rm{opt}}-{\min}_{n\in\mathcal{N}}\{D_n\})$.
\end{theorem}\begin{IEEEproof}
	Please see Appendix E.
\end{IEEEproof}
\subsection{Complexity Analysis}
	The computational complexity of the DRACS algorithm is composed of two parts, i.e., the outer layer loop for solving the fractional problem in (\ref{P3}) based on the Dinkelbach method and the inner layer loop for solving the combinatorial problem in (\ref{g_1}) based on the block coordinate descent method. In the outer layer loop, the complexity of the Dinkelbach method with $L_1$ iterations can be approximated as $\mathcal{O}\left(L_1\right)$. In the inner loop, we decompose the mixed-integer non-linear program in (\ref{g_1}) into three sub-problems in (\ref{g_1_1}), (\ref{g_1_2}), and (\ref{g_1_3}). The complexity of the block coordinate descent method with $L_2$ iterations is $\mathcal{O}\left(L_2\right)$. From Section \ref{Sec:VB}, the complexity of solving the sub-problems in (\ref{g_1_1}), (\ref{g_1_2}), and (\ref{g_1_3}) can be represented as $\mathcal{O}\left(N^2\right)$, $\mathcal{O}\left(N^2\right)$, and $\mathcal{O}\left(N^2L_3\right)$, respectively. Note that $L_3$ is the required number of iterations for solving (\ref{g_1_3}) with the Dinkelbach method. To sum up, the total computational complexity for the proposed DRACS is approximately $\mathcal{O}\left(N^2L_1L_2L_3\right)$.

\section{Experimental Results}\label{sec:Experimental Results}
\subsection{Experimental Setting}
\textcolor{black}{In our experiments, we utilize ADULT\cite{DBLP:conf/kdd/Kohavi96}, IPUMS-BR\cite{DBLP:conf/kdd/LeeK18}, MNIST, and Fashion-MNIST datasets for the i.i.d. setting to demonstrate the test accuracy.
\begin{itemize}
	\item \textbf{ADULT}. ADULT contains $45222$ individual information records of $14$ features for a binary classification task to predict whether an individual's annual income will exceed $50000$.
	\item \textbf{IPUMS-BR}. IPUMS-BR includes $38000$ individual information records of $53$ features for a binary classification task to query the range of individuals’ monthly income ($\ge300$).
	\item \textbf{MNIST}. MNIST has a training set of $60000$ $28\times28$ handwritten digits in $10$ classes (from $0$ to $9$), and a test set of $10000$ handwritten digits.
	\item \textbf{Fashion-MNIST}. Fashion-MNIST consists of $60000$ $28\times28$ grayscale images of $10$ fashion categories, along with a test set of $10000$ images.
\end{itemize}}		

\textcolor{black}{The models include squared-SVM and convolutional neural network (CNN), wherein squared-SVM model is trained on ADULT and IPUMS-BR datasets for binary classification, and CNN model is trained on MNIST and Fashion-MNIST datasets for image classification.} For the MNIST dataset, the CNN network has two 5$\times$5 convolution layers (the first with 10 channels, the second with 20, each of which is activated by ReLU, and each is followed by a 2$\times$2 max pooling layer), 2 full connected layers (the first with 320 units and the second with 50 units), and a final softmax output layer. For the Fashion-MNIST dataset, the CNN network has two 3$\times$3 convolution layers (the first with 32 channels, the second with 64, each of which is activated by ReLU, and each is followed by a 2$\times$2 max pooling layer), 3 full connected layers with 3204, 600, and 120 units, respectively, and a final softmax output layer. The loss function for squared-SVM satisfies the assumptions in Section \ref{subsec:Problem Formulation}, while CNN is non-convex and thus does not satisfy the assumptions in Section \ref{subsec:Problem Formulation}.

Besides, we set the number of clients $N=20$, and the clients are divided equally into two types. For \textbf{Type 1} clients, the size of local dataset $D_n=1000$, and the LTA energy supply $E_n^\text{sup}=600$ mW. For \textbf{Type 2} clients, $D_n=4000$, and $E_n^\text{sup}=200$ mW. \textcolor{black}{For different datasets, we set $c_n=5\times10^4$ cycles/bit and $\gamma_n=1$ Mbit for training a CNN model on the Fashion-MNIST dataset, $c_n=4\times10^4$ cycles/bit and $\gamma_n=0.8$ Mbit for training a CNN model on the MNIST dataset, $c_n=8\times10^3$ cycles/bit and $\gamma_n=0.4$ Mbit for training a squared-SVM model on the IPUMS-BR dataset, and $c_n=2\times10^3$ cycles/bit and $\gamma_n=0.1$ Mbit for training a squared-SVM model on the ADULT dataset, respectively.} \textcolor{black}{The other experimental parameters are given in Table \ref{tab:SummaryofSettings}.}
\begin{table}[!t]
	\captionsetup{labelfont={color=black,scriptsize},font={color=black,scriptsize}}		
	\caption{List of experimental parameters.}
	\label{tab:SummaryofSettings}
	\centering
	\scalebox{0.78}{
	\color{black}{\begin{tabular}{c||c|c||c|c||c|c||c|c||c|c||c}
		\hline
		\bfseries \!Parameters\!\!\! &\bfseries \!\!\!\!\!Values\!\!\!\!\!&\bfseries \!Parameters\!\!\! & \bfseries \!\!\!Values\!\!\!&\bfseries \!Parameters\!\!\!& \bfseries \!\!\!Values\!\!\!&\bfseries \!Parameters \!\!\!& \bfseries \!\!\!\!\!Values\!\!\!\!\!&\bfseries \!Parameters \!\!\!& \bfseries \!\!\!Values\!\!\!&\bfseries \!Parameters \!\!\!& \bfseries \!\!\!Values\!\!\!\\
		\hline
		$d_n$& $200$ m&$P_{n}^\text{\rm{min}}$& $23$ dBm&$h_0$& $-30$ dB&$\nu$& $2$&$f_n^{\text{\rm{max}}}$& $4$ GHz&$p_0$& $\!1\!-\!10^{-10}\!$\\
		\hline
		$d_0$& $1$ m&$P_{n}^\text{\rm{max}}$& \!$30$ dBm\!&$K$& $1$&$B$& $180$ KHz&$f_n^{\text{\rm{min}}}$& $1$ GHz&$\rho^{\text{\rm{min}}}$& $0.1$\\
		\hline	
		$\alpha$& \!$2\times10^9$\!&$\rho_n(t)$& $\sim$\text{Exp}(1)&$\beta$& $10^{\text{-3}}$&$N_0$& \!\!$-174$ dBm/Hz\!\!&$v_n$& $10^{\text{-28}}$&$\rho^{\text{\rm{max}}}$& $10$\\
		\hline						
	\end{tabular}}}
\end{table}	
\subsection{Performance of Resource Allocation and Client Scheduling}
\begin{figure}[!t]
	\centering
	\begin{minipage}[t]{0.4\textwidth}
		\includegraphics[width=\textwidth]{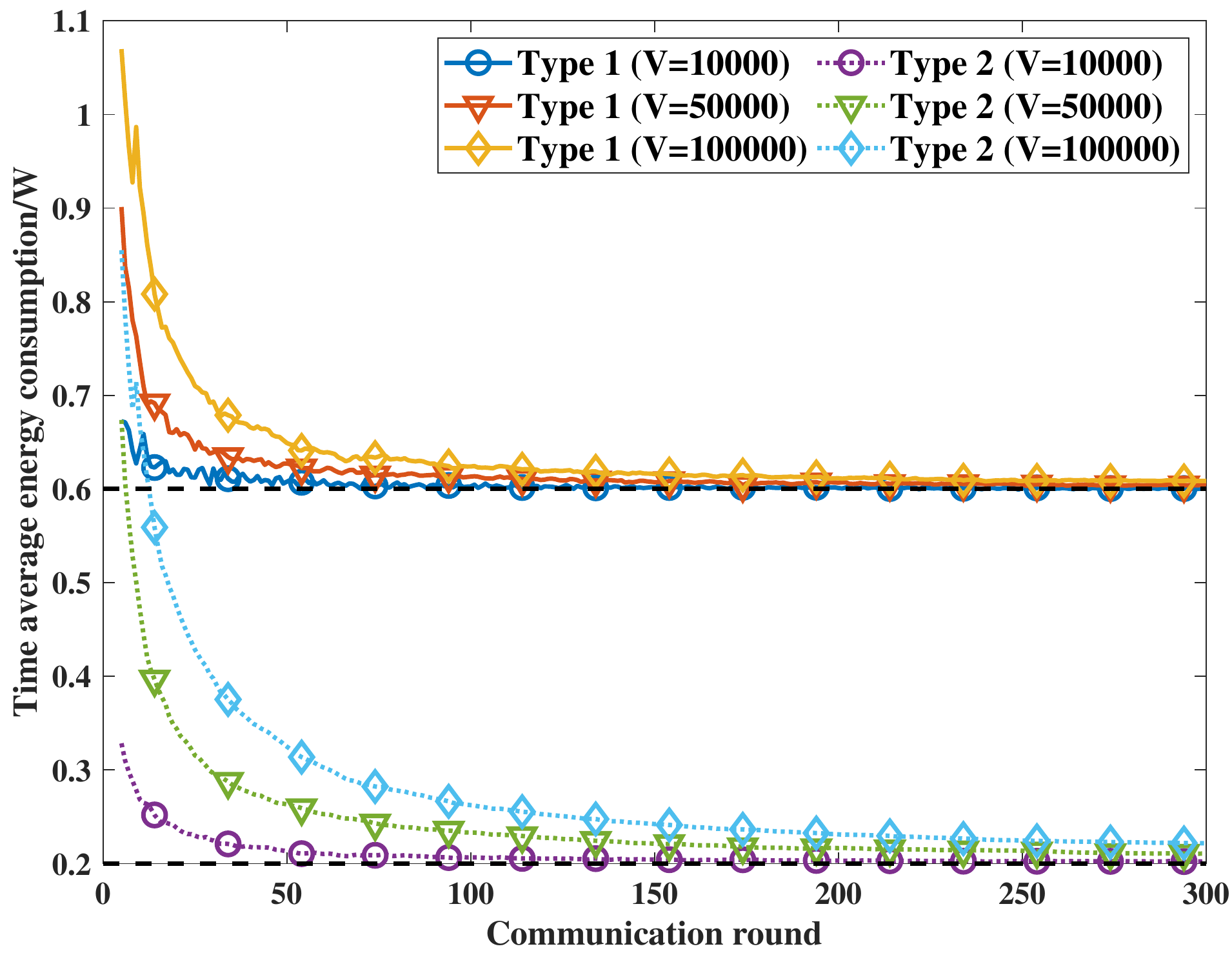}\vspace{-10pt}
		\caption{Time average energy consumption of the clients under DRACS versus $T$.}
		\label{fig:4}
	\end{minipage}
	\begin{minipage}[t]{0.4\textwidth}
		\includegraphics[width=\textwidth]{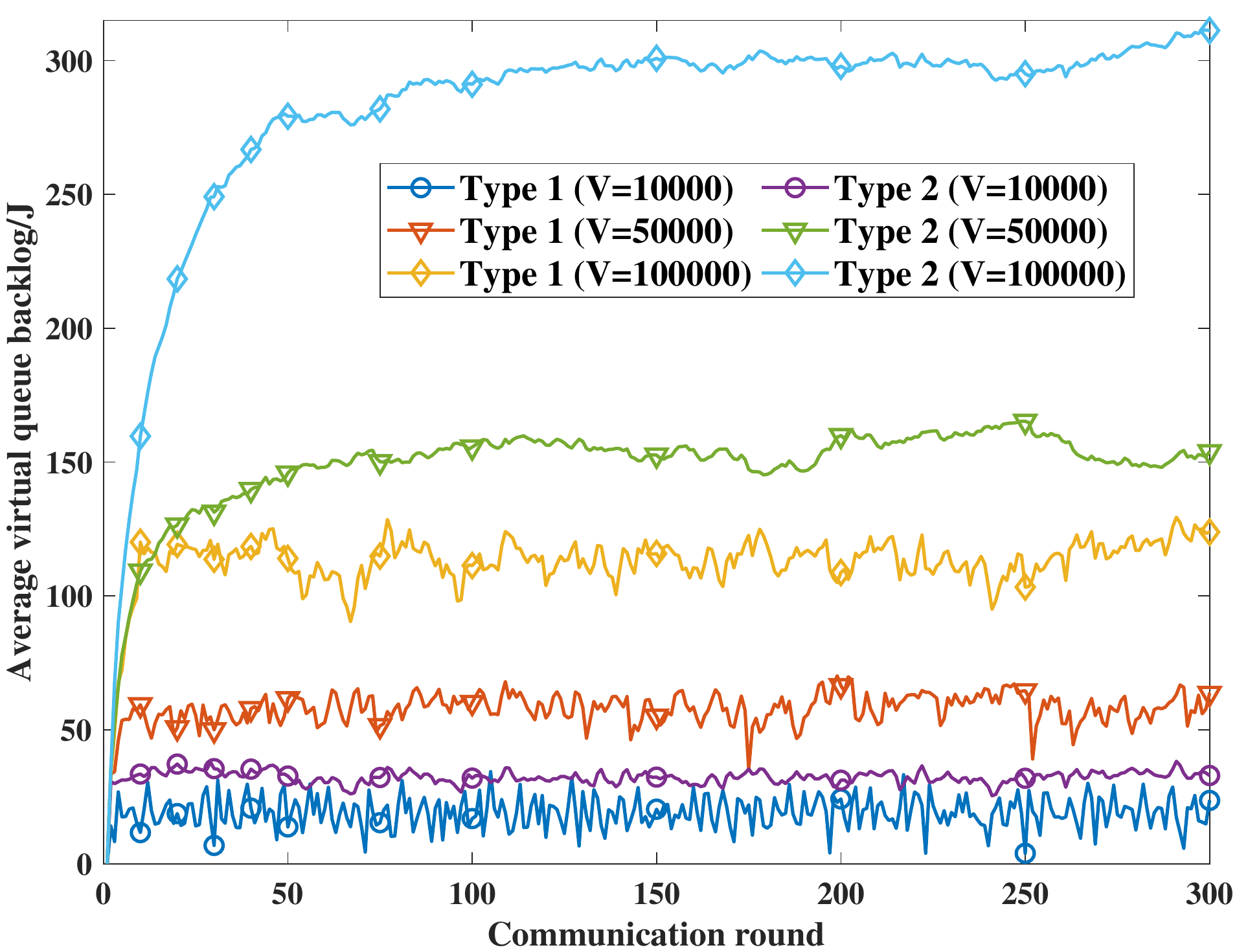}\vspace{-10pt}
		\caption{Average virtual backlogs at the clients under DRACS versus $T$.}
		\label{fig:6}
	\end{minipage}\vspace{-0.4cm}
\end{figure}
\textcolor{black}{In this subsection, we present the experimental results of the DRACS algorithm for training a CNN model on the Fashion-MNIST dataset in two parts: 1) We demonstrate the efficient allocation of energy, computation and communication resource using the proposed DRACS. 2) We compare the proposed DRACS with three benchmark client scheduling strategies in terms of the LTA training data size and energy consumption, respectively.}

Fig. \ref{fig:4} plots the LTA energy consumption of two different types of clients using DRACS with $V=10000$, $50000$, and $100000$, respectively. First, it can be observed that the LTA energy consumption of two types of clients decreases in the beginning and finally approaches the LTA energy supply as time elapses. To be specific, DRACS shows a higher LTA energy consumption than the LTA energy supply at first, but the gap between the LTA energy consumption of the clients and the LTA energy supply shrinks as $t$ increases, which is consistent with (\ref{equ3:theorem1}) in \textbf{Theorem \ref{performance analysis}}, and guarantees constraint $\textbf{C5}$ eventually. Second, we can see that the clients have the highest LTA energy consumption when $V=100000$. This is because $V$ represents how much we ignore the minimization of energy consumption and put more emphasis on the maximization of LTA training data size. Fig. \ref{fig:6} plots the time variation of average virtual backlogs at the different types of clients under DRACS with different value of $V$. It can be observed that the average virtual backlogs increase in the beginning and quickly stabilize as the time elapses.
\begin{figure}[!t]
	\centering
	\includegraphics[width=2.7in,angle=0]{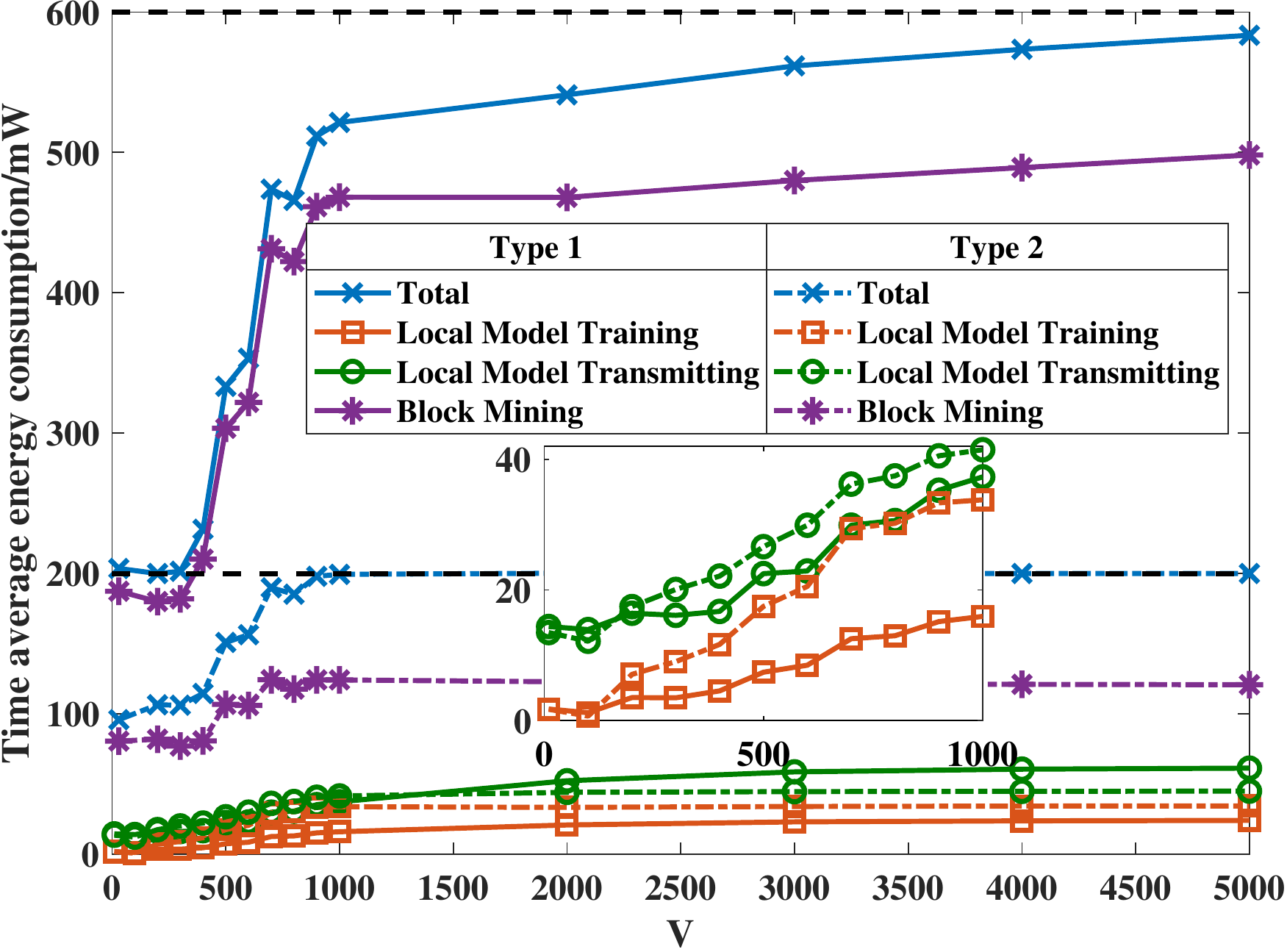}\vspace{-10pt}
	\caption{Time average energy consumption comparison between {\bf{Type 1}} and {\bf{2}} under DRACS.}\vspace{-0.6cm}
	\label{fig:13}
\end{figure}

Fig. \ref{fig:13} shows the LTA energy consumption comparison between two types of clients under DRACS over $V\in(0,5000]$. First, it is shown that the long-term time average total energy consumption at \textbf{Type 1} and \textbf{2} clients increases with $V$ when $0<V<1000$, and then stabilizes at $200$ and $600$ mW, respectively. It reveals that a relatively large value of $V$ can be adopted to fully use the supplied energy. In addition, this increasing rate first increases sharply and then slows down with $V$, which is consistent with (\ref{equ3:theorem1}) in \textbf{Theorem \ref{performance analysis}}. Second, we can see that \textbf{Type 1} clients consume less energy for local model training and much more energy for block mining than \textbf{Type 2} clients. This is due to the fact that \textbf{Type 1} clients are equipped with a smaller local dataset size than \textbf{Type 2} clients but a much higher LTA energy supply. That is, \textbf{Type 1} clients can make full use of the local energy resource to help block generation when all the clients are involved in block mining. Third, we notice that \textbf{Type 2} clients consume more energy for local model transmitting than \textbf{Type 1} clients when $0<V<1000$, and less energy when $V>1000$. This is because we emphasize more on the minimization of energy consumption and less on the maximization of training data size when minimizing the Lyapunov drift-plus-penalty ratio function $\Delta_V(t)$ in (\ref{drift-plus-penalty function'}) under a small $V\in(0,1000)$. In this case, less \textbf{Type 1} clients are involved in local model training than \textbf{Type 2} clients in order to reduce the energy consumption for local model training and transmitting, which saves energy for \textbf{Type 1} clients to mine blocks.

\begin{figure}[!t]
	\centering
	\begin{minipage}[t]{0.4\textwidth}
		\includegraphics[width=\textwidth]{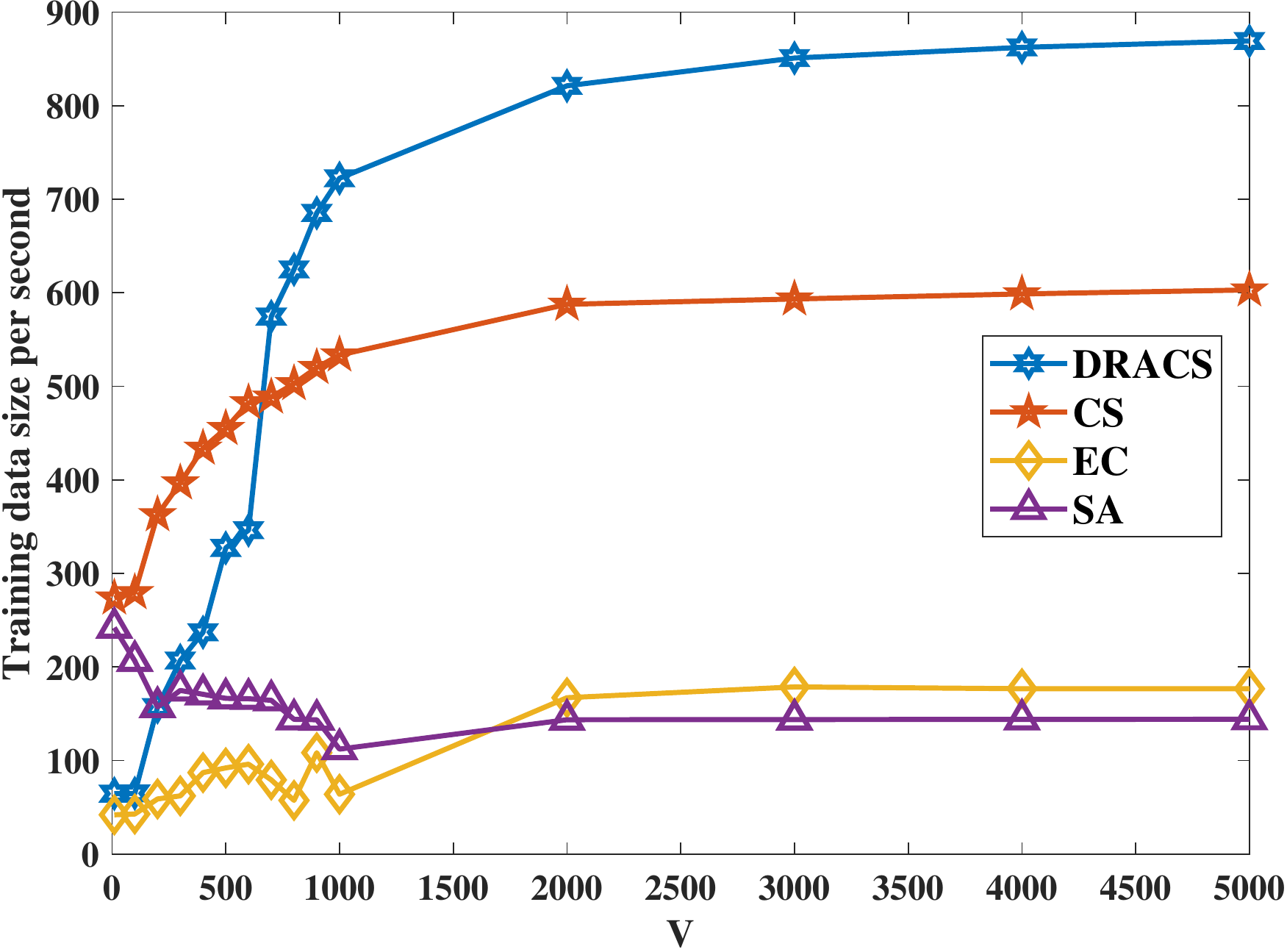}\vspace{-10pt}
		\caption{LTA total training data size comparison between DRACS and benchmark strategies.}
		\label{fig:7}
	\end{minipage}
	\begin{minipage}[t]{0.4\textwidth}
		\includegraphics[width=\textwidth]{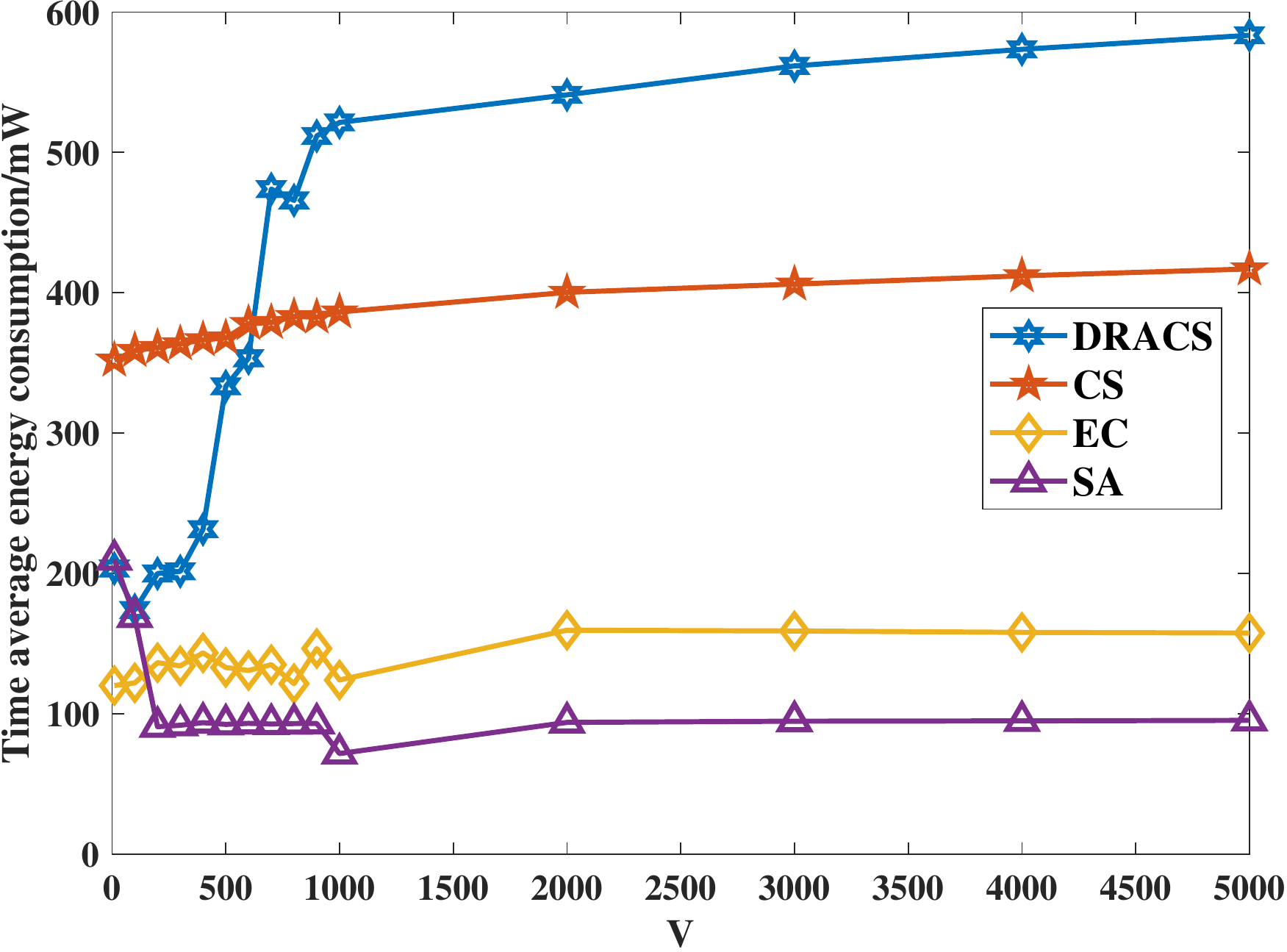}\vspace{-10pt}
		\caption{LTA energy consumption comparison at \textbf{Type 1} clients between DRACS and benchmark strategies.}
		\label{fig:9}
	\end{minipage}\vspace{-0.6cm}
\end{figure}

Fig. \ref{fig:7} and Fig. \ref{fig:9} show the LTA training data size and the LTA energy consumption at \textbf{Type 1} clients of DRACS over $V \in (0,5000]$. \textcolor{black}{For comparison purposes, we also simulate three benchmark strategies as follows: (a) client scheduling based on channel state\cite{DBLP:journals/tcom/YangLQP20} (see the line labeled with ``CS''), where the clients with high transmission rate are selected to perform local model training in each communication round; (b) client scheduling based on energy consumption (``EC''), where the clients with low LTA energy consumption are selected to perform local model training in each communication round; and (c) select all scheduling\cite{DBLP:journals/twc/LuoCWZY20} (``SA''), where all the clients are selected to perform local model training in each communication round. Note that we set the number of selected clients in each of Strategies CS and EC to be the same as for the proposed DRACS. To conduct a fair comparison, the benchmark client scheduling Strategies CS, EC, and SA first determine the client scheduling vector $\boldsymbol{i}(t)$, and then optimize the transmit power $\boldsymbol{P}(t)$, computing frequency for local model training $\boldsymbol{f}^\text{tra}(t)$, and computing frequency for block mining $\boldsymbol{f}^\text{bloc}(t)$ to maximize the training data size under the constraint of energy consumption in each communication round.} First, it can be observed in Fig. \ref{fig:7} that DRACS outperforms Strategies CS, EC, and SA. Compared with Strategies CS, EC, and SA, DRACS improves the LTA training data size effectively. This is due to the fact that DRACS relies on all information (energy consumption and channel state) as detailed in Section \ref{sec:Problem Solution} rather than partial information in Strategies CS and EC. Second, we can see that the LTA training data size of DRACS increases quickly with $V$ in the beginning and gradually stabilizes when $V \ge 3000$, which conforms to (\ref{equ2:theorem1}) in \textbf{Theorem \ref{performance analysis}}. That is, the LTA training data size of DRACS converges to the maximum utility of \textbf{P0} as $V$ increases. Third, from Fig. \ref{fig:9}, DRACS consumes more energy than Strategies CS, EC, and SA, while satisfying the energy consumption constraint $\textbf{C5}$. This reveals that DRACS can make the best use of energy in the energy-limited BFL system.
\subsection{Performance of Test Loss and Accuracy}
In this subsection, we evaluate the experimental results of test loss and accuracy performance of the proposed DRACS based on the ADULT, IPUMS-BR, MNIST and Fashion-MNIST datasets.
\begin{figure}[!t]
	\centering
	\color{black}{\subfigure[Comparison with limited time.]{\includegraphics[width=2.7in]{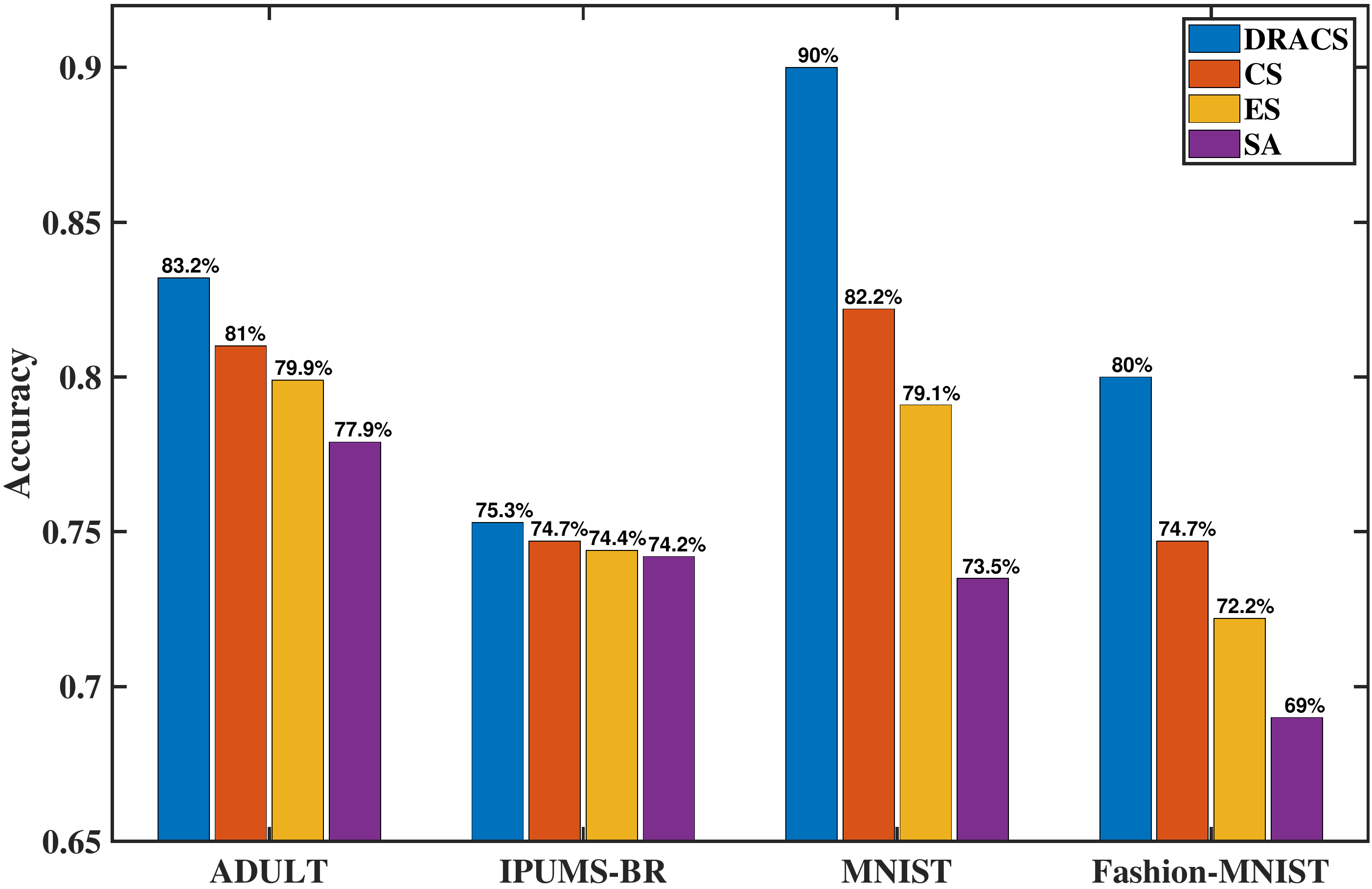}}}
	\subfigure[Comparison with limited energy consumption.]{\includegraphics[width=2.7in]{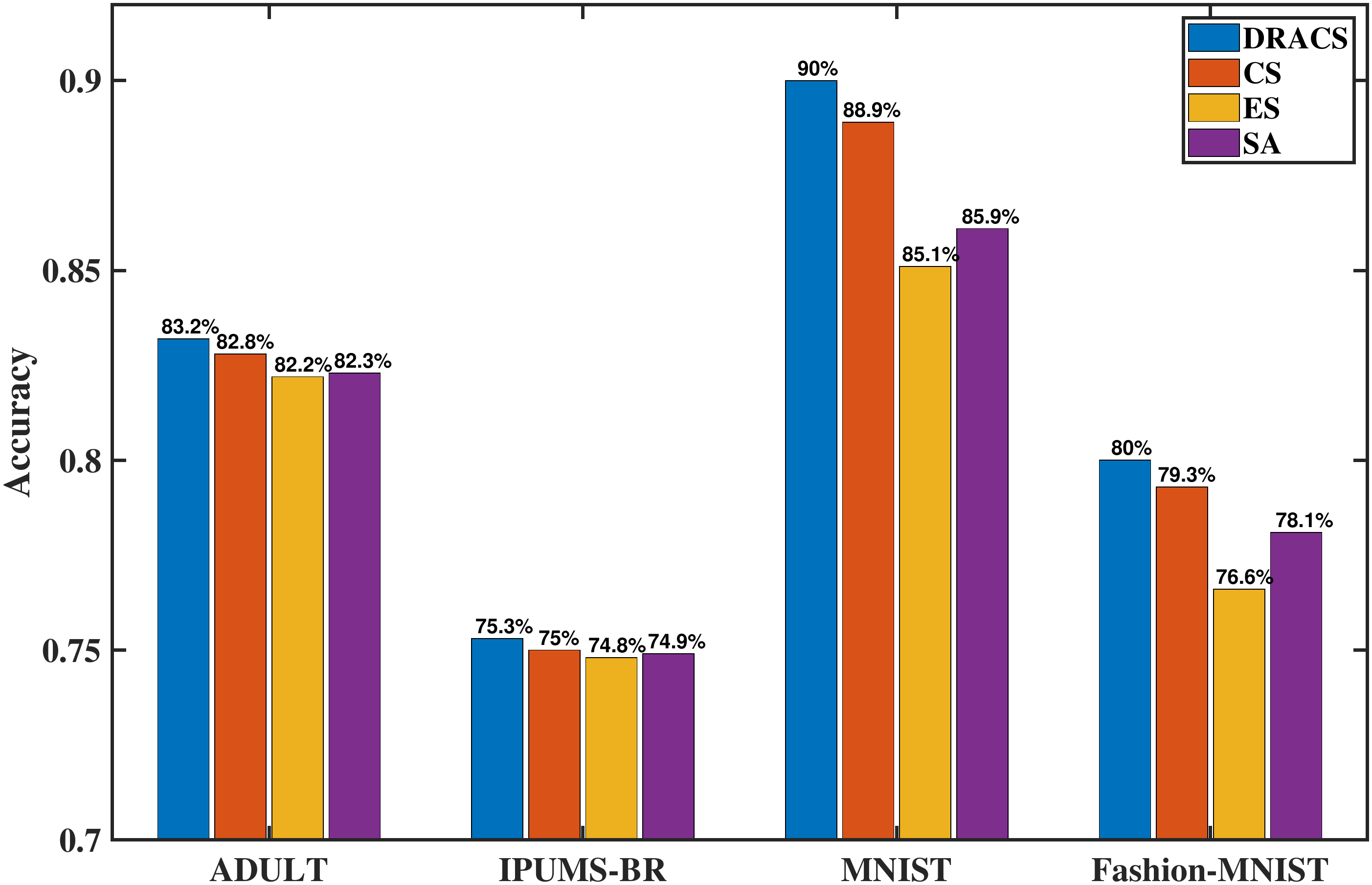}}
	\captionsetup{labelfont={color=black,scriptsize},font={color=black,scriptsize}}\vspace{-10pt}
	\caption{Accuracy comparison between DRACS and benchmark strategies with $V=30000$ on different datasets.}
	\label{fig:10}
\end{figure}
\begin{table*}[!t]
	\captionsetup{labelfont={color=black,scriptsize},font={color=black,scriptsize}}
	\caption{Test loss and communication round comparison between DRACS and benchmark strategies with $V=30000$ on different datasets.}\vspace{-10pt}
	\label{table3}
	\centering
	\color{black}{\subtable[Comparison with limited time.]{
		\scalebox{0.8}{\begin{tabular}{c||c|c|c|c|c|c|c|c}
			\hline
			& \multicolumn{2}{c|}{ADULT} & \multicolumn{2}{c|}{IPUMS-BR}& \multicolumn{2}{c|}{MNIST} & \multicolumn{2}{c}{Fashion-MNIST}\\
			\hline
			& \!\tabincell{c}{Test loss}\! &\!\tabincell{c}{Communication round}\! & \!\tabincell{c}{Test loss}\! & \!\tabincell{c}{Communication round}\! &\!\tabincell{c}{Test loss}\! &\!\tabincell{c}{Communication round}\! & \!\tabincell{c}{Test loss}\! & \!\tabincell{c}{Communication round}\!\\
			\hline\hline
			\!DRACS\! & \bf{0.395} & \bf{15} & \bf{0.569} &\bf{71} & \bf{0.348} & \bf{144} & \bf{0.552} &\bf{241}\\
			\hline
			CS & 0.420 & 9 & 0.577 &36& 0.771 & 68 & 0.705 &109\\
			\hline
			EC & 0.427 & 7 & 0.582 &30& 0.955 & 64 & 0.826 &68\\
			\hline
			SA & 0.432 & 6 & 0.586 &18& 1.289 & 43 & 0.866 &60\\
			\hline
	\end{tabular}}}}\quad\quad\quad\color{black}{\subtable[Comparison with limited energy consumption.\!\!]{
		\scalebox{0.8}{\begin{tabular}{c||c|c|c|c|c|c|c|c}
			\hline
			& \multicolumn{2}{c|}{ADULT} & \multicolumn{2}{c|}{IPUMS-BR}& \multicolumn{2}{c|}{MNIST} & \multicolumn{2}{c}{Fashion-MNIST}\\
			\hline
			& \!\tabincell{c}{Test loss}\! &\!\tabincell{c}{Communication round}\! & \!\tabincell{c}{Test loss}\! & \!\tabincell{c}{Communication round}\! &\!\tabincell{c}{Test loss}\! &\!\tabincell{c}{Communication round}\! & \!\tabincell{c}{Test loss}\! & \!\tabincell{c}{Communication round}\!\\
			\hline\hline
			\!DRACS\! & \bf{0.395} & \bf{15} & \bf{0.569} &\bf{71}& \bf{0.348} & \bf{144} & \bf{0.552} &\bf{241}\\
			\hline
			CS & 0.400 & 12 & 0.570 &61& 0.395 & 119 & 0.569 &202\\
			\hline
			EC & 0.409 & 9 & 0.575 &36& 0.551 & 91 & 0.644 &145\\
			\hline
			SA & 0.415 & 10 & 0.572 &44& 0.486 & 93 & 0.611 &155\\
			\hline
	\end{tabular}}}}\vspace{-0.6cm}
\end{table*}\textcolor{black}{Table \ref{table3} shows the comparison of the loss function and communication rounds with limited time and energy consumption between DRACS and Strategies CS, EC, and SA for the ADULT, IPUMS-BR, MNIST and Fashion-MNIST datasets, while Fig. \ref{fig:10} compares the accuracy performance.} It can be observed that, under limited time and energy consumption, the proposed DRACS achieves better learning performance than the other algorithms. This is because DRACS can execute more communication rounds (i.e., more rounds of model aggregation) than the other algorithms, by jointly optimizing communication, computation, and energy resource allocation as well as training client scheduling. The experimental results show that DRACS provides low-latency and energy-efficient resource allocation and training client scheduling protocol for both convex and non-convex loss functions.

\section{Conclusions}\label{sec:Conclusions}
In this paper, we have investigated dynamic resource management and training client scheduling in the proposed BFL network. First, we have developed a BFL framework where the functionalities of FL and blockchain are converged at the client side. Second, considering the proposed BFL framework over wireless networks, we have formulated a joint optimization problem of the training client scheduling and dynamic resource allocation to maximize the LTA training data size under the constraint of long-term time-average (LTA) energy consumption. Based on Lyapunov optimization, we have further proposed a low-complexity online DRACS algorithm to optimize the training client scheduling, transmit power, and computation frequency at the client side. Finally, experimental results have demonstrated the stability of virtual queue backlogs and corroborated the trade-off between the training data size and energy consumption. In addition, it has also been shown that DRACS can obtain higher learning accuracy than the baseline schemes under either limited training time or limited total energy supply.

Several interesting directions immediately follow from this work. First, it is of interest to analyze the theoretical convergence for non-convex loss functions. Based on the convergence analysis with non-convex assumptions on loss functions, the designs of client scheduling and resource allocation algorithms deserve further investigation. Second, to further reduce latency and energy consumption, the designs of lightweight consensus mechanisms such as Proof of Stake (PoS) for the proposed BFL framework are also of interest.


%

\ifCLASSOPTIONcaptionsoff
  \newpage
\fi

\appendices
\section{Proof of Lemma \ref{bound of Deltat}}
Recalling that $L(t)=\frac{1}{2}\sum_{n\in \mathcal{N}}Z_n(t)^2$, we have $Z_n(t+1)^2\leq Z_n(t)^2+E_{n}(t)^2+{\left(E_n^\text{sup}\right)}^2\tau(t)^2+2Z_n(t)\left(E_{n}(t)-E_n^\text{sup}\tau(t)\right)$. By moving $Z_n(t)^2$ to the left-hand side, dividing both sides by $2$, summing up the inequalities over $\emph{n}=1,\ldots,N$, and taking the conditional expectation, we have
$\Delta L\!\leq\! \sum_{n}\!\mathbb{E}\left\{\left.\!Z_n(t)(E_{n}(t)\!-\!E_n^\text{sup}\tau(t))\right\vert\boldsymbol{Z}(t)\right\}+\frac{1}{2}\sum_{n}\mathbb{E}\!\left\{\left.\!E_{n}(t)^2\!+\!{\left(E_n^\text{sup}\right)}^2\tau(t)^2\right\vert\!\boldsymbol{Z}(t)\right\}$. Given $E_{n}(t)=E_{n}^\text{tra}(t)+E_{n}^\text{up}(t)+E_n^\text{bloc}(t)$ in (\ref{energy111}), we have $\mathbb{E}\left\{\!\left.E_{n}(t)^2\right\vert\boldsymbol{Z}(t)\right\}\!\leq\!\big(v_nc_nD_n{(f_n^\text{\rm{max}})}^2\!\!-\alpha v_n\ln(1\!-$ $p_0){(f_n^\text{\rm{max}})}^3/{\sum_{n\in\mathcal{N}} f_n^\text{\rm{min}}}\big)^2+{P_n^\text{\rm{max}}\gamma_n}/{B/\log_2\left(1+\frac{P_n^\text{\rm{min}}h_0\overline{\rho_n}(d_0/d_n)^{\nu}}{(BN_0)}\right)}$. Given $\tau(t)=\max_n\{\tau_{n}^\text{tra}(t)+\tau_n^\text{up}(t)\}+\tau^\text{bloc}(t)$ in (\ref{latency}), we have $\mathbb{E}\left\{\left.{\left(E_n^\text{sup}\right)}^2\tau(t)^2\right\vert\!\boldsymbol{Z}(t)\right\}\!\leq\! {\left(E_n^\text{sup}\right)}^2\big({c_nKD_n}/{f_{n}^\text{\rm{min}}}-{\alpha\ln (1\!-\!p_0)}/\sum_{n}$ $ f_n^\text{\rm{min}}+{\gamma_n}/B/\log_2\big(1+{P_n^\text{\rm{min}}h_0\overline{\rho_n}(\frac{d_0}{d_n})^{\nu}}/{(BN_0)}\big)\big)^{2}$. Finally, by summing up $\mathbb{E}\left\{\!\left.E_{n}(t)^2\right\vert\boldsymbol{Z}(t)\right\}$ and $\mathbb{E}\left\{\left.{\left(E_n^\text{sup}\right)}^2\tau(t)^2\right\vert\boldsymbol{Z}(t)\right\}$, and dividing both sides by 2, we have $H$ in \textbf{Lemma \ref{bound of Deltat}}, which concludes proof of \textbf{Lemma \ref{bound of Deltat}}.
\section{Proof of Lemma \ref{Delta_V^*}}
First, by definition of $\Theta(t)$, we have
$\Delta_V(t)\leq\Theta(t)$. Given $0<\tau^\text{\rm{min}}\leq\tau(t)\leq \tau^\text{\rm{max}}<\infty$ in \textbf{Lemma \ref{Delta_V min max}}, we have $0\leq\big(\frac{-VD(t)+\sum_{n\in \mathcal{N}}Z_n(t)E_{n}(t)}{\tau(t)}-\Theta(t)\big)\tau(t)\leq\big(\frac{-VD(t)+\sum_{n\in \mathcal{N}}Z_n(t)E_{n}(t)}{\tau(t)}-\Theta(t)\big)\tau^\text{\rm{max}}$. By taking infimum over any $\boldsymbol{X}(t)$, we have $\!0\!\leq\!\inf_{\boldsymbol{X}(t)}\big[\big(\frac{-VD(t)+\sum_nZ_n(t)E_{n}(t)}{\tau(t)}-\Theta(t)\big)\tau(t)\big]\!\leq\!\inf_{\boldsymbol{X}(t)}\big[\big(\frac{-VD(t)+\sum_nZ_n(t)E_{n}(t)}{\tau(t)}\!-\!\Theta(t)\big)\tau^\text{\rm{max}}\big]\!=\!0$. This proves that $\inf_{\boldsymbol{X}(t)}U(t)=0$ when $\eta=\Theta(t)$.

To prove that $\inf_{\boldsymbol{X}(t)}U(t)<0$ when $\eta>\Theta(t)$, we first suppose that $\eta>\Theta(t)$. Then, we have $\inf_{\boldsymbol{X}(t)}U(t)\leq\inf_{\boldsymbol{X}(t)}\big[-VD(t)+{\sum}_{n\in \mathcal{N}}Z_n(t)E_{n}(t)-\Theta(t)\tau(t)-(\eta-\Theta(t))\tau^\text{\rm{min}}\big]=-(\eta-\Theta(t))\tau^\text{\rm{min}}<0$. To prove that $\inf_{\boldsymbol{X}(t)}U(t)>0$ when $\eta<\Theta(t)$, we first suppose that $\eta<\Theta(t)$. Then, we have $\inf_{\boldsymbol{X}(t)}U(t)\ge\inf_{\boldsymbol{X}(t)}\big[-VD(t)+{\sum}_{n\in \mathcal{N}}Z_n(t)E_{n}(t)-\Theta(t)\tau(t)+(\Theta(t)-\eta)\tau^\text{\rm{min}}\big]=(\Theta(t)-\eta)\tau^\text{\rm{min}}>0$. This concludes the proof of \textbf{Lemma \ref{Delta_V^*}}.

\textcolor{black}{
\section{The Optimal Solution of (\ref{g_1_1})}
To solve for the binary variables $\boldsymbol{i}(t)$, we first split the problem in (\ref{g_1_1}) into $N$ disjoint cases. Let $\mathcal{J}=\{1,...,N\}$ denote the index set of the $N$ cases. In each case, we assume that the $j$-th client is selected for local model training, and take the largest amount of time for local model training and transmitting, i,e., $i_j(t)=1$, and $i_{n}(t)(\tau_n^\text{tra}(t)+\tau_n^\text{up}(t))\leq i_j(t)(\tau_j^\text{tra}(t)+\tau_j^\text{up}(t))$, $\forall n\in \mathcal{N}$. Let $i_n^j(t)$ denote the optimal client scheduling vector variable of the $n$-th client in the $j$-th case. If $\tau_n^\text{tra}(t)$$+$$\tau_n^\text{up}(t)$$>\tau_j^\text{tra}(t)$$+$$\tau_j^\text{up}(t)$, the optimal client scheduling vector variable of the $n$-th client in the $j$-th case should be zero, i.e., $i_n^j(t)=0$. Otherwise, the optimal client scheduling vector variable of the $n$-th client in the $j$-th case can be determined by solving\begin{small}\begin{alignat}{1}
	\min_{i_n(t), n\in \mathcal{N}^{'}}\; &{\sum}_{n\in \mathcal{N}^{'}}i_n(t)\left\{-VD_n+Z_n(t)\bigg(v_nc_nKD_n\right.\left.\left.{\left(f_{n}^\text{tra}(t)\right)}^2+{P_n(t)\gamma_n}/{B/\log_2(1+\frac{P_n(t)h_n(t)}{(BN_0)})}\right)\right\}\label{100}\\
	&\!\!\!\!\!\!\!\!\!\!\!\!\!\!\!\!\!\text{s.t.}\quad\textbf{C1},\nonumber
\end{alignat}\end{small}where $\mathcal{N}^{'}\subseteq\mathcal{N}$ denotes the set of clients that yields $\tau_n^\text{tra}(t)+\tau_n^\text{up}(t)\leq\tau_j^\text{tra}(t)+\tau_j^\text{up}(t)$. Notably, (\ref{100}) is a standard linear program. The optimal client scheduling vector variables for the set of clients $\mathcal{N}^{'}$ in the $j$-th case is derived as\begin{small}\begin{alignat}{1}\label{101}\begin{split}i_n^j(t)=\begin{cases}
			0,\quad&\mbox{if}~-VD_n+{Z_n(t)P_n(t)\gamma_n}/{B/\log_2\left(1+\frac{P_n(t)h_n(t)}{BN_0}\right)}+Z_n(t)v_nc_nKD_n{\left(f_{n}^\text{tra}(t)\right)}^2\ge 0,\\
			1,\quad&\mbox{otherwise}.
		\end{cases}
	\end{split}
\end{alignat}\end{small}The optimal policy of training client scheduling in (\ref{101}) implies that clients with high transmission rate and low LTA energy consumption are scheduled to train their models in the current communication round.}

\textcolor{black}{
By comparing the value of $g_1(i_n^j(t))$ among the disjoint cases, we have $j_\text{opt}\!=\mathop{\arg\min}_{j\in\mathcal{J}} g_1(i_n^j(t))$. Therefore, the optimal client scheduling vector variables of (\ref{g_1_1}) is given by $i_n^*(t)={i_n^{j_\text{opt}}}(t)$.}

\textcolor{black}{
\section{Detailed Solution of (\ref{g_1_2})}
Using the approach of case analysis, we first split the optimization problem in (\ref{g_1_2}) into $R$ disjoint cases. Recall that it can be derived that $R=\sum_{n\in\mathcal{N}}i_n(t)$ clients are selected for local model training and transmitting with the optimized client scheduling vector variables $\boldsymbol{i}(t)$. Let $n_r$ denote the index of the $r$-th selected client, and ${\mathcal{N_R}=\{n_r\}}_{r=1}^{R}$ denote the set of $n_r$. In the $r$-th case, we assume that the $n_r$-th client takes the largest amount of time for local model training and transmitting, i,e., $i_{n}(t)(\tau_n^\text{tra}(t)+\tau_n^\text{up}(t))\leq\tau_{n_r}^\text{tra}(t)+\tau_{n_r}^\text{up}(t)$, $\forall n\in \mathcal{N}$. Then, we optimize the computation frequency for local model training $\boldsymbol{f}^\text{tra}(t)$ in each case, and compare the value of $g_{2}(\boldsymbol{f}^\text{tra}(t))$ among different cases.}
\textcolor{black}{
Let $f_n^{\text{tra},r}(t)$ denote the optimal computation frequency for local model training of the $n$-th client in the $r$-th case. The optimal computation frequency for local model training for the set of clients $\mathcal{N}^{r}=\{1,...,n_r-1,n_r+1,...,N\}$ in the $r$-th case can be determined by solving\begin{small}\begin{alignat}{1}
	&\min_{\boldsymbol{f}^\text{tra}(t)}\quad{\sum}_{n\in\mathcal{N}^{r}}\!i_n(t)Z_n(t)v_nc_nKD_n{\left(f_n^\text{tra}(t)\right)}^2\!\!-\!\eta{c_{n_r}KD_{n_r}}/{f_{n_r}^\text{tra}(t)}+i_{n_r}(t)Z_{n_r}(t)v_{n_r}c_{n_r}KD_{n_r}{\left(f_{n_r}^\text{tra}(t)\right)}^2\label{102}\\
	&\;\text{s.t.}\quad
	\textbf{C3}:\;f_n^\text{\rm{min}}\leq f_{n}^\text{tra}(t)\leq f_n^\text{\rm{max}}, \forall n\in \mathcal{N},\nonumber\\
	&\;\quad\quad\,\textbf{C6}:\;\frac{c_nKD_ni_n(t)}{f_{n}^\text{tra}(t)}+\frac{\gamma_ni_n(t)}{B\log_2\left(1+\frac{P_n(t)h_n(t)}{BN_0}\right)}\leq\frac{c_{n_r}KD_{n_r}}{f_{n_r}^\text{tra}(t)}i_{n_r}(t)+\frac{\gamma_{n_r}i_{n_r}(t)}{B\log_2\left(1+\frac{P_{n_r}(t)h_{n_r}(t)}{BN_0}\right)},\forall n\in \mathcal{N}.\nonumber
\end{alignat}\end{small}Obviously, the optimal computation frequency for local model training for the set of clients $\mathcal{N}^{r}$ in the $r$-th case can be derived as $f_n^{\text{tra,}r}(t)=f_n^\text{\rm{min}}$. Given the optimized computation frequency for local model training for the set of clients $\mathcal{N}^{r}$ in the $r$-th case, the optimal computation frequency of the $n_r$-th client for local model training in the $r$-th case can be determined by solving\begin{small}\begin{alignat}{1}
	&\min_{f_{n_r}^\text{tra}(t)}\quad i_{n_r}(t)Z_{n_r}(t)v_{n_r}c_{n_r}KD_{n_r}{\left(f_{n_r}^\text{tra}(t)\right)}^2-\eta{c_{n_r}KD_{n_r}}/{f_{n_r}^\text{tra}(t)}\label{103}\\
	&\text{s.t.}\quad
	\textbf{C7}:f_{n_r}^{\text{min}'}(t)\leq f_{n_r}^\text{tra}(t)\leq f_{n_r}^\text{\rm{max}},\nonumber
\end{alignat}\end{small}where $\small{ f_{n_r}^{\text{min}'}\!(t)\!=\!\max\!\bigg\{\!i_{n_r}\!(t)c_{n_r}KD_{n_r}\!\bigg(\!\max\limits_{n\in\mathcal{N}^{r}}\!\!\bigg\{\!\frac{i_n(t)c_nKD_n}{f_{n}^\text{tra}(t)}+\frac{i_n(t)\gamma_n}{B\log_2(1\!+\!\frac{P_n(t)h_n(t)}{BN_0})}\bigg\}-\frac{i_{n_r}(t)\gamma_{n_r}}{B\log_2(1\!+\!\frac{P_{n_r}(t)h_{n_r}(t)}{BN_0})}\!\bigg)^{\!\!-1}\!,}$ $\small{f_{n_r}^\text{\rm{min}}\bigg\}}$. Notably, (\ref{103}) is a continuous derivable function. The optimal computation frequency of the $n_r$-th client for local model training in the $r$-th case is given by\begin{small}\begin{equation}\label{104}
	\begin{split}
		f_{n_r}^{\text{tra,}r}(t)=
		\begin{cases}
			\!f_{n_r}^{\text{min}'}(t),\,\mbox{if}~\eta\ge0,\,\mbox{else if}~\left({-\eta c_{n_r}KD_{n_r}}/({2i_{n_r}(t)Z_{n_r}(t)v_{n_r}c_{n_r}KD_{n_r}})\right)^{\frac{1}{3}}\leq f_{n_r}^{\text{min}'}(t),\\
			\!f_{n_r}^\text{\rm{max}},\,\mbox{if}~\;\!\eta<0\,\,\text{and}\left({-\eta c_{n_r}KD_{n_r}}/{(2i_{n_r}(t)Z_{n_r}(t)v_{n_r}c_{n_r}KD_{n_r})}\right)^{\frac{1}{3}}\ge f_{n_r}^\text{\rm{max}},\\
			\!\left({-\eta c_{n_r}KD_{n_r}}/{(2i_{n_r}(t)Z_{n_r}(t)v_{n_r}c_{n_r}KD_{n_r})}\right)^\frac{1}{3},\,\mbox{otherwise}.
		\end{cases}
	\end{split}
\end{equation}\end{small}}
\textcolor{black}{
Thus, given the optimized computation frequency for local model training $f_{n}^{\text{tra,}r}(t)$ in each case, we have $r_\text{opt}=\mathop{\arg\min}_r g_{2}(f_{n}^{\text{tra,}r}(t))$. Therefore, the optimal computation frequency for local model training is given by ${f_n^\text{tra}}^*(t)=f_n^{\text{tra}{,r_{\rm{opt}}}}(t)$.}

\section{Proof of Theorem \ref{performance analysis}}
Before we show the main proof of \textbf{Theorem \ref{performance analysis}}, we first give \textbf{Lemma \ref{iid alorithm}} and \textbf{\ref{Optimality over i.i.d. algorithms}} which will be used to compare the LTA training data size and LTA energy consumption of any possible i.i.d. policy with the maximum LTA training data size $\phi^\text{opt}$ and the LTA energy supply $E_n^\text{sup}$.
\begin{definition}
	A policy is i.i.d., if it takes a control action $\boldsymbol{X}(t)$ independently and probabilistically according to a single distribution in each communication round.	
\end{definition}
Let $\boldsymbol{\Gamma}$ denote the set of expectations of time averages $\overline{D}$, $\overline{\tau}$, and $\{\overline{E_{n}}\}_{n\in\mathcal{N}}$ under all possible i.i.d. policies, where $\overline{D}=\lim_{T\to\infty}\frac{\sum_{t=1}^TD(t)}{T}$, $\overline{\tau}=\lim_{T\to\infty}\frac{\sum_{t=1}^T\tau(t)}{T}$, and $\overline{E_{n}}=\lim_{T\to\infty}\frac{\sum_{t=1}^TE_{n}(t)}{T}$. Note that the set of expectation of time averages $\boldsymbol{\Gamma}$ is bounded and convex.
\begin{lemma}\label{iid alorithm}
	Under any possible policy $\pi$ that meets all constraints of \rm{\textbf{P0}} ($\textbf{C1} \sim \textbf{C5}$), we have
	\begin{alignat}{1}
		\!\!\mathbb{E}\left\{\left.\left[D(t),E_1(t),...,E_N(t),\tau(t)\right]\right\vert\pi\right\}\!\in\! \boldsymbol{\Gamma},\; \frac{1}{T}{\sum}_{t=0}^{T-1}\mathbb{E}\left\{\left.\left[D(t),E_1(t),...,E_N(t),\tau(t)\right]\right\vert\pi\right\}\in \boldsymbol{\Gamma}\label{iid alorithm2}.\!\!\!\!\!
	\end{alignat}
\end{lemma}
(\ref{iid alorithm2}) holds by considering the policy $\pi$ taking a control action independently and probabilistically according to a distribution in each communication round as one that is from an i.i.d. policy. Recall that $\phi^\text{opt}$ is defined as the maximum utility of \textbf{P0} over all possible policies. Consider a policy $\pi^0$ that meets all constraints of \textbf{P0} ($\textbf{C1}\sim\textbf{C5}$) and it yields
\begin{small}\begin{alignat}{1}
	\lim_{T\to\infty}\sup \left[\frac{\frac{1}{T}\sum_{t=0}^{T-1}\mathbb{E}\{D(t)\vert\pi^0\}}{\frac{1}{T}\sum_{t=0}^{T-1}\mathbb{E}\{\tau(t)\vert\pi^0\}}\right]\ge \phi^\text{opt}-\frac{\delta}{2},\quad\lim_{T\to\infty}\sup \left[\frac{\frac{1}{T}\sum_{t=0}^{T-1}\mathbb{E}\{E_{n}(t)\vert\pi^0\}}{\frac{1}{T}\sum_{t=0}^{T-1}\mathbb{E}\{\tau(t)\vert\pi^0\}}\right]\leq E_n^\text{sup}.
\end{alignat}\end{small}It follows that for a finite integer $T_0$, we have
\begin{small}\begin{alignat}{1}
	\frac{\frac{1}{T_0}\sum_{t=0}^{T_0-1}\mathbb{E}\{D(t)\vert\pi^0\}}{\frac{1}{T_0}\sum_{t=0}^{T_0-1}\mathbb{E}\{\tau(t)\vert\pi^0\}}\ge \phi^\text{opt}-\delta,\quad\frac{\frac{1}{T_0}\sum_{t=0}^{T_0-1}\mathbb{E}\{E_{n}(t)\vert\pi^0\}}{\frac{1}{T_0}\sum_{t=0}^{T_0-1}\mathbb{E}\{\tau(t)\vert\pi^0\}}\leq E_n^\text{sup}+\delta.\label{euq9:theorem}
\end{alignat}\end{small}By \textbf{Lemma \ref{iid alorithm}}, there exists an i.i.d. policy $\pi'$ such that
\begin{alignat}{1}\label{euq7:theorem}
	&\frac{1}{T_0}{\sum}_{t=0}^{T_0-1}\mathbb{E}\left\{\left.\left[D(t),E_1(t),...,E_N(t),\tau(t)\right]\right\vert\pi^0\right\}=\mathbb{E}\left\{\left.\left[D(t),E_1(t),...,E_N(t),\tau(t)\right]\right\vert\pi'\right\}.
\end{alignat}Plugging (\ref{euq7:theorem}) into the both equations in (\ref{euq9:theorem}), and it yields
\begin{alignat}{1}
	{\mathbb{E}\left\{D(t)\vert\pi'\right\}}/{\mathbb{E}\left\{\tau(t)\vert\pi'\right\}}\ge \phi^\text{opt}-\delta,\quad{\mathbb{E}\{E_{n}(t)\vert\pi'\}}/{\mathbb{E}\left\{\tau(t)\vert\pi'\right\}}\leq E_n^\text{sup}+\delta.\label{euq11:theorem}
\end{alignat}Multiplying both sides of the equations in (\ref{euq11:theorem}) by $\mathbb{E}\{\tau(t)\vert\pi'\}$, and it proves \textbf{Lemma \ref{Optimality over i.i.d. algorithms}} below.
\begin{lemma}\label{Optimality over i.i.d. algorithms}
	For any $\delta>0$, there exists an i.i.d. policy $\pi'$ that satisfies
	\begin{alignat}{1}
		\mathbb{E}\{D(t)\vert\pi'\}\ge\mathbb{E}\{\tau(t)\vert\pi'\}(\phi^\text{\rm{opt}}-\delta),\quad\mathbb{E}\{E_{n}(t)\vert\pi'\}\leq\mathbb{E}\{\tau(t)\vert\pi'\}(E_n^\text{\rm{sup}}+\delta), \forall n\in\mathcal{N}.\label{equ2:lemma6}
	\end{alignat}
\end{lemma}

Second, to prove (\ref{equ2:theorem1}), from (\ref{upper bound of Delta(t)}), we have
\begin{alignat}{1}\label{euq12:theorem}
	&\Delta L-V\mathbb{E}\{D(t)\vert\boldsymbol{Z}(t)\}\leq H+{\sum}_{n\in \mathcal{N}}\mathbb{E}\{-VD(t)+Z_n(t)(E_{n}(t)-E_n^\text{sup}\tau(t))\vert\boldsymbol{Z}(t)\}.
\end{alignat}Substituting (\ref{C-additive approximation}) into (\ref{euq12:theorem}), it yields $\small\Delta L-V\mathbb{E}\{D(t)\vert\boldsymbol{Z}(t)\}\leq H-\!\sum_{n\in \mathcal{N}}Z_n(t)E_n^\text{sup}\mathbb{E}\{\tau(t)\vert\boldsymbol{Z}(t)\}+\mathbb{E}\{\tau(t)\vert\boldsymbol{Z}(t)\}\left(C\!+\!\!\sum_{n\in \mathcal{N}}\!\frac{\mathbb{E}\left\{-VD(t)\!+\!Z_n(t)E_{n}(t)\vert\boldsymbol{Z}(t),\pi'\right\}}{\mathbb{E}\left\{\tau(t)\vert\boldsymbol{Z}(t),\pi'\right\}}\right)$, where $\pi'$ is any possible i.i.d. policy. Plugging the both equations in (\ref{equ2:lemma6}) into the right-hand-side, and letting $\delta\to 0$, we have
\begin{alignat}{1}\label{equ4:theorem1}
	\!\Delta L\!-\!V\mathbb{E}\{D(t)\vert\boldsymbol{Z}(t)\}\!\leq\! H\!+\!(C\!-\!V\phi^\text{opt})\mathbb{E}\{\tau(t)\vert\boldsymbol{Z}(t)\}.
\end{alignat}By summing up the equalities in (\ref{equ4:theorem1}) over $t=0,1,...,T$, and dividing both sides by $\overline{\tau}(T)$ and $T$, we have
$\frac{\overline{D}(T)}{\overline{\tau}(T)}\ge \phi^\text{opt}-\frac{H/\overline{\tau}(T)+C}{V}-\frac{\mathbb{E}\{L(T)-L(0)\}}{\overline{\tau}(T)VT}$, where $\overline{D}(T)=\frac{\sum_{t=0}^{T-1}\mathbb{E}\{D(t)\}}{T}$, and $\overline{\tau}(T)=\frac{\sum_{t=0}^{T-1}\mathbb{E}\{\tau(t)\}}{T}$. Note that $\mathbb{E}\{L(t)\}<\infty$. Therefore, we have (\ref{equ2:theorem1}) with $\delta\to 0$.

Third, to prove (\ref{equ3:theorem1}), it can be derived from (\ref{equ4:theorem1}) that $\Delta L\leq H+\tau^\text{\rm{max}}(C-V\phi^\text{opt})+V{\sum}_{n\in \mathcal{N}}D_n$, where the inequality holds since $\mathbb{E}\{D(t)\vert\boldsymbol{Z}(t)\}\leq{\sum}_{n\in \mathcal{N}}D_n$, and $\mathbb{E}\{\tau(t)\vert\boldsymbol{Z}(t)\}\leq \tau^\text{\rm{max}}$. Let $G_1=2(H+\tau^\text{\rm{max}}C)$, and $G_2=2(\tau^\text{\rm{max}}\phi^\text{opt}-{\sum}_{n\in \mathcal{N}}D_n)$, we have $\Delta L\leq \frac{G_1-VG_2}{2}$. By summing up $\Delta L\leq \frac{G_1-VG_2}{2}$ over $t=0,1,2,...,T-1$, taking expectations, dividing both sides by $T$, and recalling that $L(t)=\frac{1}{2}\sum_{n\in \mathcal{N}}Z_n(t)^2$, we have ${\sum}_{n\in \mathcal{N}}\frac{\mathbb{E}\{Z_n(T)^2\}}{T}\leq G_1-VG_2+{\sum}_{n\in \mathcal{N}}\frac{\mathbb{E}\{Z_n(0)^2\}}{T}$. Thus, for each client $n\in\mathcal{N}$, we have $\frac{\mathbb{E}\{Z_n(T)^2\}}{T}\leq G_1-VG_2+{\sum}_{n\in \mathcal{N}}\frac{\mathbb{E}\{Z_n(0)^2\}}{T}$. By dividing both sides by $T$, and squaring both sides, we have $\frac{\mathbb{E}\{Z_n(T)\}}{T}\leq\sqrt{ \frac{G_1-VG_2}{T}+{\sum}_{n\in \mathcal{N}}\frac{\mathbb{E}\{Z_n(0)^2\}}{T^2}}$, since Jensen's inequality shows that $\mathbb{E}\{Z_n(T)\}^2\leq\mathbb{E}\{Z_n(T)^2\}$. From (\ref{vitual queues}), we have
\begin{alignat}{1}\label{equ17:theorem1}
	Z_n(t+1)\ge Z_n(t)+E_{n}(t)-E_n^\text{sup}\tau(t).
\end{alignat}By summing up (\ref{equ17:theorem1}) over $t=0,1,2,...,T-1$, taking expectations, dividing both sides by $T$, and noting that $\mathbb{E}\{Z_n(0)\}<\infty$, we have $\frac{\mathbb{E}\{Z_n(T)\}}{T}\ge \overline{E_{n}}(T)-E_n^\text{sup}\overline{\tau}(T)$, where $\overline{E_{n}}(T)=\frac{\sum_{t=0}^{T-1}\mathbb{E}\{E_{n}(t)\}}{T}$. Finally, we have
\begin{small}\begin{alignat}{1}
		\frac{\overline{E_{n}}(T)}{\overline{\tau}(T)}&\leq E_n^\text{sup}+\frac{\mathbb{E}\{Z_n(T)\}}{\overline{\tau}(T)T}\leq E_n^\text{sup}+\frac{\mathbb{E}\{Z_n(T)\}}{\tau^\text{\rm{min}}T}\leq E_n^\text{sup}+\frac{1}{\tau^\text{\rm{min}}}\sqrt{ \frac{G_1-VG_2}{T}+\sum_{n\in \mathcal{N}}\frac{\mathbb{E}\{Z_n(0)^2\}}{T^2}}.
\end{alignat}\end{small}This concludes the proof of \textbf{Theorem \ref{performance analysis}}.

\bibliographystyle{IEEEtran}
\bibliography{references}

\end{document}